\documentclass[11pt]{article}
\usepackage{tabularx}
\usepackage{cancel,slashed}
\usepackage{bbm}
\usepackage{amssymb}
\usepackage{amsfonts}
\usepackage{graphicx}
\usepackage{cite}
\usepackage{physics}
\usepackage{amsmath}
\usepackage{tikz}
\usepackage{float}
\restylefloat{table}
\usepackage[normalem]{ulem}
\usepackage{tensor}

\usepackage{ dsfont }
\usepackage{latexsym}
\usepackage{color}
\usepackage{cancel,slashed}
\usepackage[hyperfootnotes=false,linktocpage]{hyperref}
\usepackage{color}
\usepackage{transparent}
\usepackage[hang,flushmargin]{footmisc}

\newcommand{\jq}[1]{{\bf[{\color{red}JQ:} #1]}}

\newcommand{\bk}[1]{{\color{black} #1}}

\usepackage{blkarray}
\usepackage{multirow}
\usepackage{empheq}

\usepackage{comment}

\newcommand{\bmat}{\left(\begin{array}}
\newcommand{\emat}{\end{array}\right)}

\def\1{{\bf 1}}
\def\2{{\bf 2}}
\def\3{{\bf 3}}
\def\4{{\bf 4}}
\def\6{{\bf 6}}

\def\targ#1#2{\genfrac{[}{]}{0pt}{}{#1}{#2}}
\def\targ2#1#2{\genfrac{}{}{0pt}{}{#1}{#2}}

\definecolor{mygr}{rgb}{0,0.6,0}
\definecolor{mygrey}{rgb}{0,0.1,0.2}
\definecolor{myblue}{rgb}{0,0.5,0.9}
\definecolor{myblue2}{rgb}{0,0.5,0.5}
\definecolor{myblue3}{rgb}{0,0.7,0.9}
\definecolor{myblue4}{rgb}{0,0.6,0.6}
\definecolor{myorange}{rgb}{1,0.5,0}
\definecolor{mypurple}{rgb}{0.6,0,1}
\definecolor{mygolden}{rgb}{1,0.8,0.2}
\definecolor{mycyan}{rgb}{0,1,1}
\definecolor{mymagenta}{rgb}{1,0,1}
\definecolor{mykiwi}{rgb}{0.8,1,0.5}
\definecolor{mybrown}{cmyk}{0.14, 0.42, 0.56, 0.2}
\definecolor{myturq}{cmyk}{0.99, 0, 0.2, 0.4}
\definecolor{myaubergine2}{cmyk}{0.4, 0.5, 0, 0.1}
\definecolor{myaubergine}{cmyk}{0.6,0.85,0,0}
\definecolor{CycleGreen}{cmyk}{0.52,0,1,0}
\definecolor{CycleBrown}{cmyk}{0, 0.4, 0.9, 0.2}

\DeclareFontFamily{U}{rcjhbltx}{}
\DeclareFontShape{U}{rcjhbltx}{m}{n}{<->rcjhbltx}{}
\DeclareSymbolFont{hebrewletters}{U}{rcjhbltx}{m}{n}

\DeclareMathSymbol{\lamed}{\mathord}{hebrewletters}{108}
\DeclareMathSymbol{\mem}{\mathord}{hebrewletters}{109}
\DeclareMathSymbol{\ayin}{\mathord}{hebrewletters}{96}
\DeclareMathSymbol{\tsadi}{\mathord}{hebrewletters}{118}
\DeclareMathSymbol{\qof}{\mathord}{hebrewletters}{113}
\DeclareMathSymbol{\resh}{\mathord}{hebrewletters}{114}
\DeclareMathSymbol{\pe}{\mathord}{hebrewletters}{112}
\DeclareMathSymbol{\pesofit}{\mathord}{hebrewletters}{80}
\DeclareMathSymbol{\samekh}{\mathord}{hebrewletters}{115}
\DeclareMathSymbol{\tav}{\mathord}{hebrewletters}{116}
\DeclareMathSymbol{\vav}{\mathord}{hebrewletters}{119}
\DeclareMathSymbol{\het}{\mathord}{hebrewletters}{120}
\DeclareMathSymbol{\yod}{\mathord}{hebrewletters}{121}
\DeclareMathSymbol{\zayin}{\mathord}{hebrewletters}{122}
\DeclareMathSymbol{\alephdot}{\mathord}{hebrewletters}{128}
\DeclareMathSymbol{\tsadisofit}{\mathord}{hebrewletters}{90}
\DeclareMathSymbol{\shin}{\mathord}{hebrewletters}{152}

\def\be{\begin{equation}}
\def\ee{\end{equation}}
\def\bea{\begin{eqnarray}}
\def\eea{\end{eqnarray}}
\def\bes{\begin{subequations}}
\def\ees{\end{subequations}}

\usepackage{multicol}
\usepackage{float}
\usepackage{caption}
\newenvironment{eqn*}{\begin{equation*}\begin{aligned}}{\end{aligned}\end{equation*}\noindent}

\makeatletter
\newsavebox\myboxA
\newsavebox\myboxB
\newlength\mylenA

\newcommand*\xoverline[2][0.75]{%
\sbox{\myboxA}{$\m@th#2$}%
\setbox\myboxB\null
\ht\myboxB=\ht\myboxA%
\dp\myboxB=\dp\myboxA%
\wd\myboxB=#1\wd\myboxA
\sbox\myboxB{$\m@th\overline{\copy\myboxB}$}
\setlength\mylenA{\the\wd\myboxA}
\addtolength\mylenA{-\the\wd\myboxB}%
\ifdim\wd\myboxB<\wd\myboxA%
   \rlap{\hskip 0.5\mylenA\usebox\myboxB}{\usebox\myboxA}%
\else
    \hskip -0.5\mylenA\rlap{\usebox\myboxA}{\hskip 0.5\mylenA\usebox\myboxB}%
\fi}
\makeatother

\begin{document}

\begin{flushright}
\hfill{}
\end{flushright}
\vspace{0.5cm}

	\begin{center}        
		\huge Regge growth of isolated massive spin-2 particles \\and the Swampland
	\end{center}
	
	\vspace{0.7cm}
	\begin{center}        
		{\large  Suman Kundu$^1$,\;\; Eran Palti$^2$,\;\; Joan Quirant$^2$, }
	\end{center}
	
	\vspace{0.15cm}
	\begin{center}  
	\emph{${}^1$Department of Particle Physics and Astrophysics,\\
	The Weizmann Institute of Science, Rehovot 76100, Israel}\\[.2cm]
		\emph{${}^2$Department of Physics, Ben-Gurion University of the Negev,\\ Be'er-Sheva 84105, Israel}\\[.3cm]
		\emph{}\\[.2cm]
		e-mails:  \tt suman.kundu@weizmann.ac.il, \; palti@bgu.ac.il, \;joanq@post.bgu.ac.il

	\end{center}
	
	\vspace{1cm}
	
	
	\begin{abstract}
	\noindent  
	We consider an effective theory with a single massive spin-2 particle and a gap to the cutoff. We couple the spin-2 particle to gravity, and to other lower-spin fields, and study the growth of scattering amplitudes of the particle in the Regge regime: where $s$ is much larger than $t$ and also any mass scales in the effective theory, but still much lower than the cutoff scale of the theory and therefore any further massive spin-2 particles. We include in the effective theory all possible operators, with an arbitrary, but finite, number of derivatives. We prove that the scattering amplitude grows strictly faster than $s^2$ in any such theory. Such fast growth goes against expected bounds on Regge growth. We therefore find further evidence for the Swampland spin-2 conjecture: that a theory with an isolated massive spin-2 particle, coupled to gravity, is in the Swampland. 
	\end{abstract}
	
	\thispagestyle{empty}
	\clearpage

\tableofcontents

\section{Introduction}
\label{s:intro}

The motivation for this work is the question of whether a theory with a single massive spin-2 particle coupled to gravity, which has a gap to any other spin-2 or higher-spin particles, is consistent. Our analysis will crucially rely on the assumption that the theory includes gravity, and the consistency of the theory will be tested due to this coupling. In this sense, we are motivated to understand if a theory with an isolated massive spin-2 particle  is in the Swampland (see \cite{Palti:2019pca,vanBeest:2021lhn} for reviews). Swampland constraints which relate the existence of a massive spin-2 particle to the cutoff of the effective theory were proposed in \cite{Klaewer:2018yxi}. Indeed, the statement was made that a massive spin-2 particle with a parametric gap to the cutoff is not consistent. We find further evidence for this statement in this work. We discuss more details on the connection to the spin-2 conjecture in section \ref{sec:relsp2c}.

\begin{figure}[h]
\centering
\includegraphics[width=14cm]{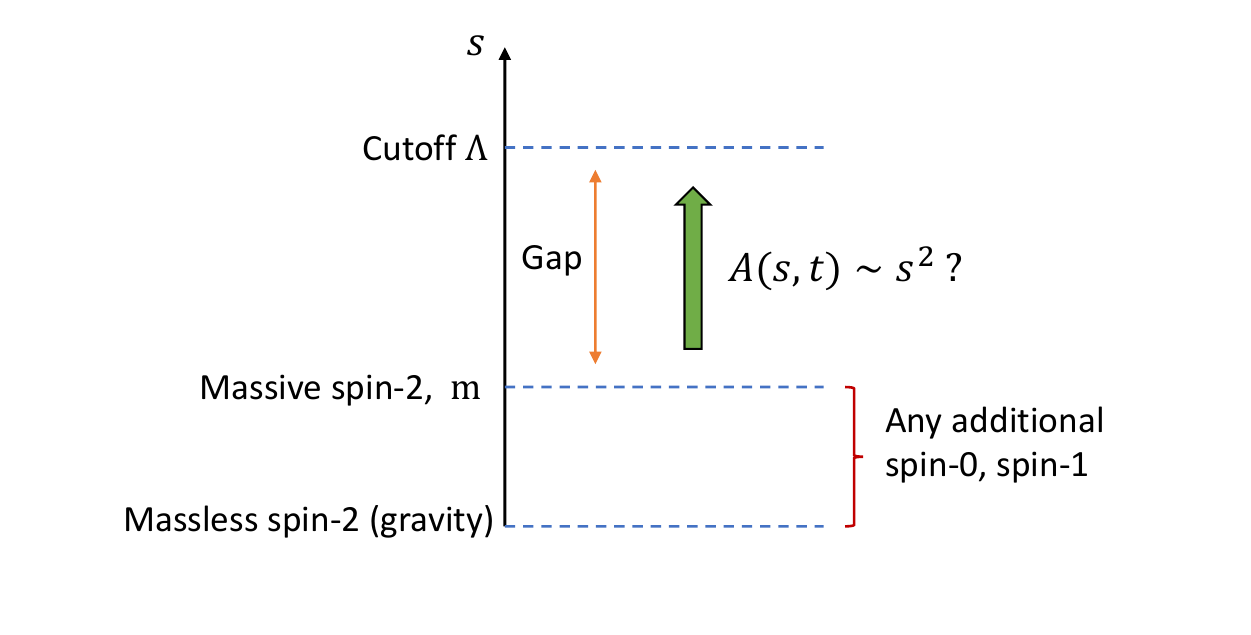}
\caption{Figure showing the setup of the effective theory that we consider. There is a parametric gap between the mass of the massive spin-2 particle $m$ and the cutoff of the effective theory $\Lambda$. The Regge growth regime we consider is between $m$ and $\Lambda$. As well as gravity, there can be other massive or massless spin-0 or spin-1 particles in the theory, though we demand that their masses are also below the Regge regime. The question we address is whether the amplitude Regge growth can be no faster than $s^2$.}
\label{fig:gap}
\end{figure}

The approach we take to this question in this work is the behaviour of two-to-two classical scattering amplitude ${\cal A}$ of the massive spin-2 particle. In particular, we consider how the amplitude grows with the centre-of-mass energy parameter $s$, in the regime where $s$ is much larger than the momentum exchange parameter $t$ and the mass of the spin-2 particle $m$, but still much smaller than the cutoff scale $\Lambda$ of the theory. In such a regime there is a leading power of $s$ that will dominate the amplitude, and we aim to constrain the theory by demanding that this power cannot be too large, more precisely, cannot be larger than two:
\begin{equation}
	\label{crgc}
	s\, \partial_s \log {\cal A}(s,t) \leq 2  \quad\mathrm{for}\quad \Lambda \gg \sqrt{s}\gg m\;,\sqrt{|t|} \;\;.
\end{equation}
The question we address is whether there exists a theory with an isolated massive spin-2 particle which can satisfy the constraint (\ref{crgc}). We will allow the theory also to have further (a finite number of) massless or massive spin-0 and spin-1 particles, but with a mass below the Regge regime defined above. This is illustrated in figure \ref{fig:gap}. 
		
The constraint (\ref{crgc}) is a version of the Classical Regge Growth conjecture (CRG) proposed in \cite{Chowdhury:2019kaq}. The original conjecture, as formulated in \cite{Chowdhury:2019kaq}, is the statement that the classical (tree-level) S-matrix of any consistent theory cannot grow faster than $s^2$ in the Regge limit, that is for $s \rightarrow \infty$ at fixed $t$. This is not precisely (\ref{crgc}), but it certainly inspires it. 

The bound (\ref{crgc}) is a local bound in the language of \cite{Haring:2022cyf}. Some strong results on such a bound were developed in \cite{Haring:2022cyf}, for example, showing that it holds in five dimensions or higher for scattering of scalar particles. We refer to \cite{Haring:2022cyf} for references on the topic, to \cite{deRham:2022hpx} for a recent review, and to \cite{Chandorkar:2021viw,deRham:2022gfe,Noumi:2022wwf,Hamada:2023cyt} for other recent work.\footnote{In this work we will consider only flat space scattering amplitudes. In AdS, there is some evidence that the Regge growth bound is related to the Chaos bound in the dual CFT \cite{Camanho:2014apa,Maldacena:2015waa,Chowdhury:2019kaq,Chakraborty:2020rxf}.} 

The theories that we consider, below the cutoff $\Lambda$, are the most general local effective theories with a {\it finite number of derivatives}. So we allow for arbitrary higher-derivative operators, but do not account for the possibility of an infinite number of {\it correlated} higher derivative operators that can resum into non-analyticities. Such series must correspond to integrating out a particle and are controlled by the mass scale of the particle that has been integrated out. If the particle is spin-0 or spin-1, we include this possibility explicitly by allowing for such states in the effective theory below the cutoff $\Lambda$. If the particle that has been integrated out has spin-2 or higher, then by construction it must have a mass above the cutoff $\Lambda$. Since we are working in the regime $\Lambda \gg \sqrt{s}$, we can reliably neglect such possibilities. In other words, we are interested in theories where one massive spin-2 particle is isolated from other massive spin-2 (or higher) particles, and that includes their effects through an infinite number of correlated higher-derivative operators.

The primary quantity that we are interested in is the growth of the scattering amplitude with $s$. There are many other ways to constrain theories with massive (and massless) spin-2 particles. A small selection of relevant papers follows. Perhaps most similar in nature to our analysis are the papers \cite{Bonifacio:2018vzv, Bonifacio:2018aon, Bonifacio:2019mgk,Bonifacio:2019ioc}, on which we rely heavily. These papers studied the total energy dependence of scattering amplitudes in the type of theories we are considering. They then used this to bound the cutoff scale through unitarity, or to constrain the spectrum of particles. In particular, in \cite{Bonifacio:2019ioc} the constraint on the energy growth was shown to lead to a beautiful bound on the spectrum of Kaluza-Klein states in compactifications of higher dimensional pure gravity. There are similar papers which study constraints imposed by superluminality \cite{Camanho:2014apa,Hinterbichler:2017qyt, Bonifacio:2017nnt}. There are also many papers studying theories of purely massive spin-2 states, so without the massless spin-2 gravity present. We refer to \cite{deRham:2014zqa,Hinterbichler:2011tt} for reviews, and to \cite{Bonifacio:2016wcb,deRham:2017xox,Bellazzini:2017fep,deRham:2018qqo,Alberte:2019xfh,Wang:2020xlt, Bellazzini:2023nqj} for some relevant work.

\subsection*{Summary of results}

It is simple to summarise our results: we find that there are no possible theories, so any values of the couplings of any operators, which have tree-level scattering that can satisfy the Regge growth bound (\ref{crgc}).

As stated, this result holds up to the assumption that there are no relevant infinite correlated series of higher dimension operators, that is, that there is a gap to any poles of further massive spin-2 particles. 

We also restrict to four dimensions: $d=4$. This makes the computation more tractable as there are identities which reduce the number of operators. We do not expect our results to change in higher dimensions.

The implications of our results for theories with isolated massive spin-2 particles, coupled to gravity, depend on how strongly one expects the classical Regge growth bound (\ref{crgc}) to hold. There is very strong evidence for this, but to the best of our knowledge, it remains to be proven.

\subsection{Relation to the Spin-2 conjecture}
\label{sec:relsp2c}

In this section we discuss some aspects of the spin-2 conjecture proposed in \cite{Klaewer:2018yxi}. The conjecture states that in a theory with gravity and a massive spin-2 particle of mass $m$, there is a bound on the cutoff $\Lambda$ of the theory
\be
\Lambda \sim \frac{m M_p}{M_w} \;.
\ee  
Here $M_p$ is the Planck scale and $M_w$ is a mass scale which sets the interactions of the massive spin-2 particle. So it is a scale which appears in the coupling of the field, $h_{\mu\nu}$, to the tensor current $T^{\mu\nu}$  which defines its interactions
\be
\frac{1}{M_w} h_{\mu\nu} T^{\mu\nu} \;.
\ee
If $h_{\mu\nu}$ was the graviton, then $M_w$ would be $M_p$, and $T^{\mu\nu}$ would be the energy-momentum tensor.

The scale $M_w$ is somewhat subtle, because the interaction strength may vary for different fields. It is therefore natural to suggest that the cutoff is set by the weakest interaction scale (largest value of $M_w$), otherwise the proposal is not well-posed and leads to different cutoffs associated with different fields. A universal interaction which is always present is gravity, and this is always controlled by the mass scale $M_p$. This means that if the massive spin-2 field is coupled to gravity we should take $M_w \sim M_p$. This is then the natural application of the spin-2 conjecture to our setup, which gives the proposed constraint of $\Lambda \sim m$, so that there cannot be a parametric gap from the mass of the spin-2 to an infinite tower of states.

\section{Classifying scattering amplitudes }
\label{sec:ampli}

This section aims to present all the ingredients needed to compute the most general $2\rightarrow 2$ tree-level scattering amplitude of a massive spin-2 particle. To do so, we will consider that the massive spin-2 particle can couple to a (massive or massless) scalar particle, a massive spin-1 particle and a massless spin-2 particle. Fermions do not need to be introduced since by momentum conservation they cannot be exchanged between bosons. Massless spin-1 particles can also be ignored, since its coupling with two massive spin-2 particles is not allowed by gauge symmetry.

Tree-level scattering amplitudes can be computed directly, without reference to specific Lagrangian terms, in a model-independent way. One can use the fact that the result has to satisfy Lorentz invariance, crossing symmetry, locality and unitarity to chart all the possible contributions. This can be done using on-shell methods, as explained for instance in \cite{Costa:2011mg}. The basic idea behind on-shell amplitudes is that they are invariant under field redefinitions and integration by parts in the Lagrangian, making the classification of the vertices easier.   Most of the computations  of this part were already developed in \cite{Bonifacio:2018vzv, Bonifacio:2018aon, Bonifacio:2019mgk}, in a different context.\footnote{The authors of these papers were interested in the high energy limit, $\{s\rightarrow +\infty\, ,\,  t\rightarrow -\infty\, ,\,  s/t \rightarrow \text{fixed}\}$, of the same scattering.} For that reason, we will only explain the basic steps in the main text, relegating some details to appendices \ref{ap:kin}-\ref{ap:fpver} and referring the reader to the aforementioned references for a more detailed discussion.

Let us start by fixing the notation. A particle $i$ has momentum $p_i$, spin $l_i$ and mass $m_i$. We denote its polarisation tensor by $\epsilon_i$. This tensor is symmetric, traceless and satisfies $p_i^\mu \epsilon_{i\mu}=0$. Momentum conservation in our conventions reads $p_1+p_2+p_3=0$. Formally, when constructing the amplitudes, we will write the polarisation matrices  as a product of vectors, $\epsilon_i=\epsilon_{i\mu\nu}\equiv\epsilon_{i\mu} \epsilon_{i\nu}$, with these vectors satisfying $\epsilon_{i\mu} \epsilon_i^{\mu}=0$, $\epsilon_{i\mu} p_i^\mu=0$. This does not mean that we are assuming the physical polarisations matrices to have  rank one:  it is only a way in which one can keep track of the contractions more easily. At any point of the computation one can always go back to the $\epsilon_{\mu\nu}$ formalism. In the new language,  gauge invariance of the massless spin-1 and spin-2 particles means invariance under $\epsilon_{i\mu}\rightarrow \epsilon_{i\mu}+ \alpha p_{i\mu}$, being $\alpha$ an arbitrary constant. We will denote $\epsilon_{i\mu}p_j^\mu \equiv A_{ij}$, $\epsilon_{i\mu} \epsilon_j^{\mu}\equiv B_{ij}=B_{ji}$. All the fields are taken to be canonically normalised, with mass dimension one. Parity-odd interactions always involve contractions with the Levi-Civita tensor $\varepsilon$, $\varepsilon\left(p_i,p_j,\epsilon_k,\epsilon_l\right)\equiv \varepsilon^{\mu\nu\alpha\beta}\,p_{i\mu}\, p_{j\nu} \,\epsilon_{k\alpha}\,\epsilon_{l\beta} $.  Finally, we denote by $\mathcal{M}_{a,b,c}\left(m_a,m_b,m_c\right)$ a three-point interaction of particles   with spin $a$, $b$, $c$ and masses $m_a$, $m_b$, $m_c$, where the particle $c$ will be the one exchanged.

\begin{figure}[h]
\centering
\includegraphics[width=12cm]{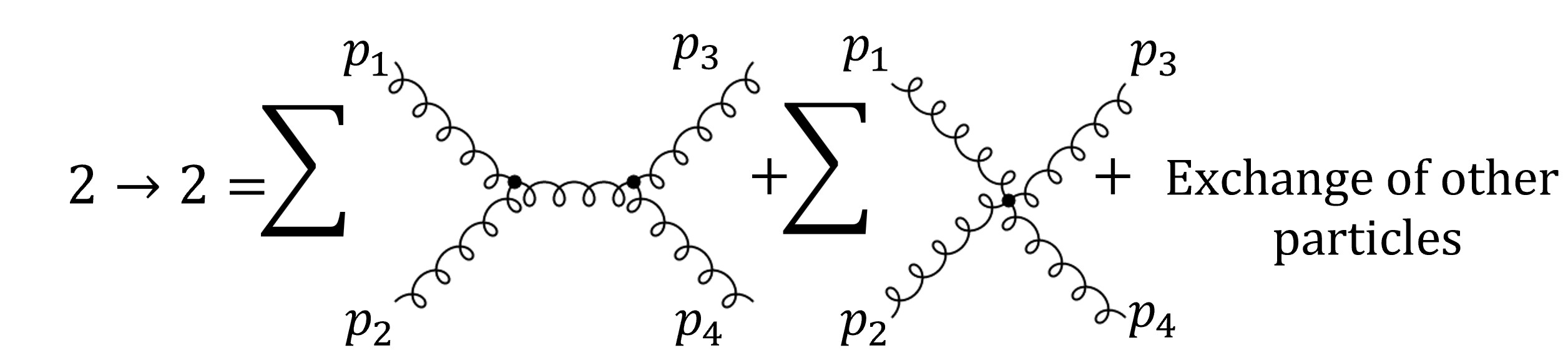}
\caption{Contributions to the tree-level $2\rightarrow2$ scattering of any self-interacting particle.}
\label{figure:contandex}
\end{figure}

Two sources of diagrams contribute to any $2\rightarrow 2$ tree-level scattering amplitude: exchange diagrams and contact terms, $\mathcal{A}_{2\rightarrow 2}=A^{\rm contact}+A^{\rm exhcange}$, pictorially represented in figure \ref{figure:contandex}. We discuss them case by case.

\subsection{On shell three-point interactions}
\label{sec:threepoint}

\begin{figure}[h!]
\centering
\includegraphics[width=15cm]{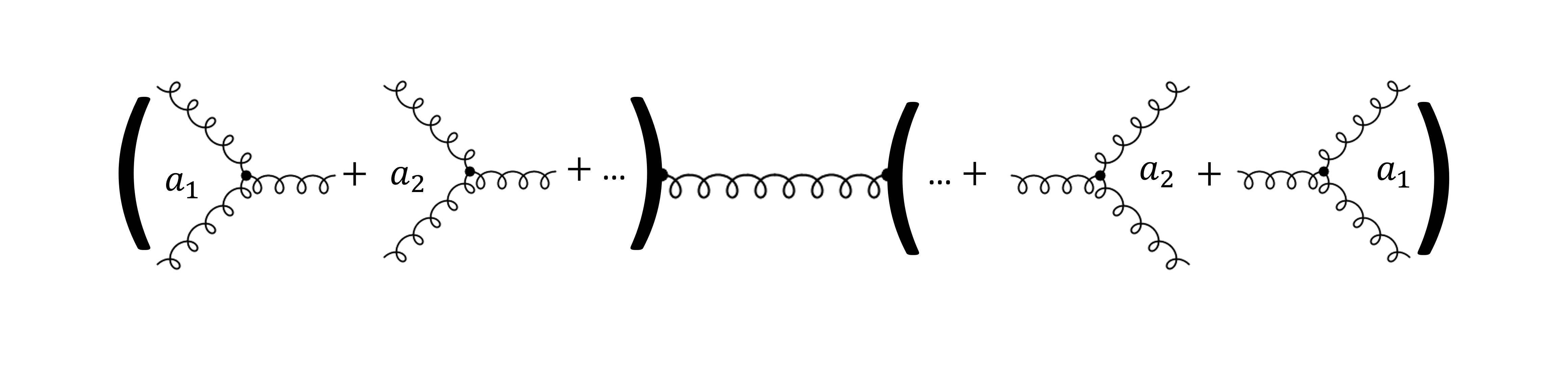}
\caption{The exchange amplitude can be computed using on-shell three-point vertices. Each vertex is assigned a coefficient $a_i$.}
\label{figure:exex}
\end{figure}

The exchange diagrams can be computed in two steps. First, we need to list all the possible on-shell three-point vertices between the two massive spin-2 particles and the exchanged particle. Then, we take two sets of these vertices, and connect them through the corresponding propagator, see figure \ref{figure:exex}.

This reduces the problem to find all possible on-shell 3-point interactions allowed by Lorentz invariance, crossing symmetry, unitarity, locality and gauge invariance. We discuss how to do this in appendix \ref{ap:onshell}, presenting directly the results in the next subsections.

\subsubsection{Massive-2, massive-2, massive-2}

In this section, we are looking at three-point couplings between three identical massive spin-2 particles of mass $m$. There are in general ten contributions, five of them are parity-even and the other five are parity-odd \cite{Bonifacio:2017iry, Bonifacio:2018vzv}. One of them (parity-even) comes from a renormalizable piece in the Lagrangian and the rest from non-renormalizable pieces. In four dimensions, $d=4$, this list can be reduced  by taking into account that  any set of more than four vectors cannot be linearly independent. We discuss these redundancies in appendix   \ref{ap:pd3pfunc}, showing directly in table \ref{MMM2}  the independent functions.
\begin{table}[H]
\centering
\begin{tabular}{c }
$\mathcal{M}_{2,2,2}\left(m,m,m\right)$   \\ \hline
 $\mu B_{12}B_{13}B_{23}\equiv \mathcal{M}_1$	\\
  $\frac{1}{\Lambda_2}\left( B_{13}^2A_{21}^2+B_{23}^2A_{12}^2+B_{12}^2A_{31}^2\right)\equiv \mathcal{M}_2$	\\
 $\frac{1}{\Lambda_3}\left( B_{12}B_{23}A_{12}A_{32}+B_{12}B_{13}A_{21}A_{31}+B_{13}B_{23}A_{12}A_{21}\right)\equiv \mathcal{M}_3$	 \\
$\frac{1}{\Lambda_4^5}A_{12}^2A_{23}^2A_{31}^2\equiv \mathcal{M}_4$	 \\
$ \frac{1}{\Lambda_5}\left( B_{13}\, B_{23}\, \varepsilon \left( p_1,p_2,\epsilon_1,\epsilon_2\right)-B_{12}\, B_{23}\, \varepsilon \left( p_1,p_2,\epsilon_1,\epsilon_3\right)+B_{12}\, B_{13}\, \varepsilon \left( p_1,p_2,\epsilon_2,\epsilon_3\right)\right)\equiv \mathcal{M}_5$\\
$\frac{1}{\Lambda_{6}^5}\left( A_{12}A_{23}A_{31}\left(A_{31}\,\varepsilon \left( p_1,p_2,\epsilon_1,\epsilon_2\right)-A_{23}\,\varepsilon \left( p_1,p_2,\epsilon_1,\epsilon_3\right)+A_{12}\,\varepsilon \left( p_1,p_2,\epsilon_2,\epsilon_3\right)\right)\right)\equiv  \mathcal{M}_{6}$\\

	\end{tabular}
\caption{Independent three-point amplitudes for three identical massive  spin 2-particles with mass $m$ in $d=4$.  } \label{MMM2}
\end{table}

Here the spin-2 particle is taken to have mass $m$, and the coefficients $\mu$ and $\Lambda_i$ have energy dimension one. Generally, we assume, here and in the next sections, that the scale $\Lambda_i$ at which every term becomes relevant can be different. A table with the complete list of the ten contributions is  shown in appendix \ref{ap:pd3pfunc}. A Lagrangian basis generating the  parity-even elements (the full list, not only the ones presented in table \ref{MMM2})  can be found in appendix  \ref{ap:submasmasimasi}.\footnote{This basis is not one-to-one since, as explained above, Lagrangians related by field redefinitions or total derivatives give rise to the same on-shell vertices}

The total cubic vertex we will be considering for this interaction is therefore (restricting ourselves to four dimensions):
\begin{align}
\boxed{V_{2,2,2}(m,m,m)\equiv a_1 \mathcal{M}_1+a_2 \mathcal{M}_2+a_3 \mathcal{M}_3+a_4\mathcal{M}_5+a_5\mathcal{M}_5+a_6\mathcal{M}_6} \;.
\label{eq:vermassive}
\end{align}
with $a_i$ being arbitrary coefficients.

\subsubsection{Massive-2, massive-2, massless-2}
\label{sec:masimasimass}

In this case we start with six  parity-even and  nine parity-odd three-point operators at high enough dimension. We can exploit again the fact that we are working in $d=4$ to write some of them as linear combinations of the others. A list with the fifteen interactions is given in appendix \ref{ap:pd3pfunc}, in table \ref{MM02a}, whereas below, in table \ref{MM02}, we only write the nine (five even and four odd) linearly independent contributions \cite{Bonifacio:2017nnt}.
\begin{table}[H]
\centering
\begin{tabular}{c}
$\mathcal{G}_{2,2,2}\left(m,m,0\right)$   \\  \hline
$\frac{1}{M_p} A_{31}^2 B_{12}^2\equiv \mathcal{G}_1$ \\
$\frac{1}{M_p} A_{31} B_{12} \left(A_{12} B_{23}+A_{23} B_{13}\right)\equiv \mathcal{G}_2$  \\
$\frac{1}{M_p}\left(A_{23} B_{13}+A_{12} B_{23}\right){}^2\equiv \mathcal{G}_3$	\\
$\frac{1}{M_p \hat\Lambda_4^2}A_{12} A_{21} A_{31}^2 B_{12}\equiv \mathcal{G}_4$\\
$\frac{1}{M_p \hat{\Lambda}_5^4}A_{12}^2 A_{23}^2 A_{31}^2\equiv \mathcal{G}_5$	 \\
$\frac{1}{M_p}\left(B_{23}A_{12}+B_{13}A_{23}\right)\,\varepsilon\left(p_3 ,\epsilon_1 , \epsilon_2 , \epsilon_3\right)\equiv \mathcal{G}_6$\\
$\frac{1}{M_p}B_{12}A_{31}\,\varepsilon\left(p_3 ,\epsilon_1 , \epsilon_2 , \epsilon_3\right)\equiv \mathcal{G}_7$\\
$\frac{1}{M_p\hat{\Lambda}_8^2}A_{12}A_{23}A_{31}\,\varepsilon\left(p_3 ,\epsilon_1, \epsilon_2 , \epsilon_3\right)\equiv \mathcal{G}_8$\\
$\frac{1}{M_p\hat{\Lambda}_{9}^4}A_{12}A_{23}A_{31}^2\,\varepsilon\left(p_1 , p_2 ,\epsilon_1\, \epsilon_2\right)\equiv \mathcal{G}_{9}$
\end{tabular}
\caption{Basis of independent three-point amplitudes for   two (identical) massive spin-2 particles, one massless  spin 2-particle in four dimensions. } \label{MM02}
\end{table}

The massless spin-2 particle is the  graviton, and so each vertex is suppressed with a factor of the Planck mass $M_p$. The $\hat\Lambda_i$ have units of energy and a Lagrangian basis for the parity-even interactions is given in appendix \ref{ap:submasmasimass}. The total cubic vertex for this interaction is therefore:
\begin{align}
\boxed{V_{2,2,2}(m,m,0)\equiv g_1 \mathcal{G}_1+g_2 \mathcal{G}_2+g_3 \mathcal{G}_3+g_4\mathcal{G}_4+g_5\mathcal{G}_5+g_6\mathcal{G}_6+g_7\mathcal{G}_7+g_8\mathcal{G}_8+g_{9}\mathcal{G}_{9}}\, .
\label{eq:vermassles}
\end{align}
Here the $g_i$ are arbitrary coefficients. Moreover, as discussed in \cite{Bonifacio:2018aon,Bonifacio:2019mgk} and the references therein,  gauge invariance requires
\begin{align}
\label{eq:gaugeinva}
2g_1=g_2\, .
\end{align}
This relation can be derived by studying the Compton scattering of a massive and a massless spin-2 particle, which involves both the massive-2, massive-2, massless-2 and the massless-2, massless-2, massless-2 vertices (the latter coming from the usual graviton self-interaction cubic vertex in the Einstein-Hilbert term).

\subsubsection{Massive-2, massive-2, massive-1}

Exploiting one more time the fact that we are interested in the case $d=4$, in table \ref{M2M2M1} we present the (independent) three-point vertices we will be considering, agreeing with \cite{Bonifacio:2018aon}. We relegate the full list of interactions for any dimension to appendix \ref{ap:pd3pfunc}, table \ref{M2M2M1a}.
\begin{table}[H]
\centering
\begin{tabular}{c}
$\mathcal{M}_{2,2,1}\left(m,m,m_1\right)$  \\ \hline
$\left( A_{12} B_{12} B_{23}+A_{21} B_{12} B_{13}\right)\equiv \mathcal{K}_1$\\
 $\frac{1}{\tilde{\Lambda}_2^2}\left( A_{21} A_{12}^2 B_{23}+A_{21}^2 A_{12} B_{13}\right)\equiv \mathcal{K}_2$\\
$B_{12}\left(\varepsilon\left(p_1,\epsilon_1,\epsilon_2,\epsilon_3\right)-\varepsilon\left(p_2,\epsilon_1,\epsilon_2,\epsilon_3\right)\right)\equiv \mathcal{K}_3$\\
$\frac{1}{\tilde{\Lambda}_4^2}\, \varepsilon\left(p_1,p_2,\epsilon_1,\epsilon_2\right)\left(B_{23}A_{12}+B_{13}A_{21}\right)\equiv \mathcal{K}_4$\\
\end{tabular}
\caption{Four-dimensional three-point amplitudes for two identical  massive  spin-2 particles and one massive spin-1 particle} \label{M2M2M1}
\end{table}
Here $\tilde{\Lambda}_i$  is the scale of suppression of the vertex $\mathcal{K}_i$ and a parity-even Lagrangian  basis is detailed in appendix \ref{ap:submasmasimasi1}. The total cubic vertex for this case is therefore:
\begin{align}
\boxed{V_{2,2,1}(m,m,m_1)\equiv k_1 \mathcal{K}_1+k_2 \mathcal{K}_2+k_3 \mathcal{K}_3+k_4 \mathcal{K}_4}\, .
\label{eq:verspin1}
\end{align}
The $k_i$ are arbitrary coefficients.

\subsubsection{Massive-2, massive-2, scalar}

Finally, regarding the coupling with a scalar field $\phi$, our computations match  the ones of  \cite{Bonifacio:2019mgk}:

\begin{table}[H]
\centering
\begin{tabular}{c}
$\mathcal{M}_{2,2,0}\left(m,m,M\right)$	   \\ \hline
	$m_s B_{12}^2=\mathcal{S}_1$ \\
		$\frac{1}{\bar\Lambda_2} A_{12} A_{21} B_{12}=\mathcal{S}_2$ \\

	$\frac{1}{\bar\Lambda_3^3} A_{12}^2 A_{21}^2=\mathcal{S}_3$ \\
$\frac{1}{\bar\Lambda_4} B_{12}\, \varepsilon\left(p_1,p_2,\epsilon_1,\epsilon_2\right)=\mathcal{S}_4$	\\
$\frac{1}{\bar\Lambda_5^3}  A_{12}A_{23}\, \varepsilon\left(p_1,p_2,\epsilon_1,\epsilon_2\right)=\mathcal{S}_5$	\\
\end{tabular}
\caption{All possible on-shell three-point amplitudes for two  massive  spin 2-particles and one scalar particle} \label{M2M2M0}
\end{table}
Here $\bar\Lambda_i$ suppress the non-renormalizable operators, $m_s$ is some interaction scale, the mass $M$ of the scalar can be $M\geq 0$ and a basis of Lagrangian terms for the parity-even amplitudes is derived in appendix \ref{ap:submasmasscalar}. The total vertex for this interaction is:
\begin{align}
\boxed{V_{2,2,0}(m,m,M)\equiv s_1 \mathcal{S}_1+s_2 \mathcal{S}_2+s_3 \mathcal{S}_3+s_4 \mathcal{S}_4+s_5 \mathcal{S}_5}\, .
\label{eq:versscalar}
\end{align}
The $s_i$ are arbitrary coefficients.

\subsection{Exchange amplitude}
\label{sec:exch}

As explained previously, the contribution of an exchanged particle $k$ to the on-shell $2\rightarrow 2$ scattering of a massive spin-2 particle can be computed as follows. Firstly, we  take two sets of vertices $V_{2,2,k}\left(m,m,m_k\right)$ defined in equations \eqref{eq:vermassive}, \eqref{eq:vermassles}, \eqref{eq:verspin1} and \eqref{eq:versscalar}. Secondly, we ``remove" the particle $k$ of the  vertices and connect them through the corresponding propagator, given explicitly in appendix \ref{ap:kin}.
\begin{figure}[h]
\centering
\includegraphics[width=16cm]{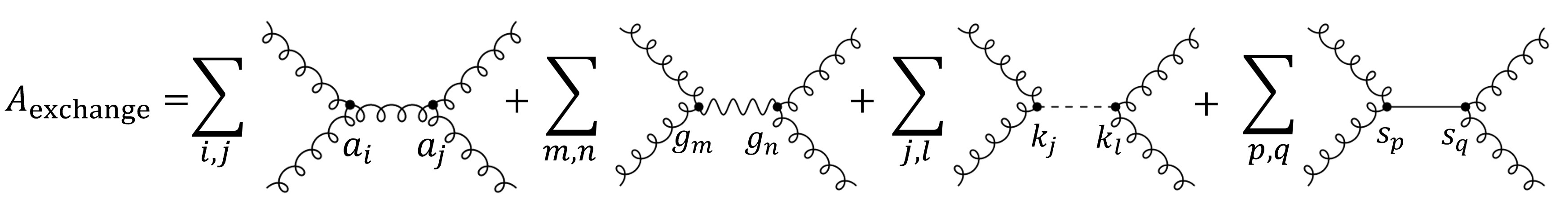}
\caption{To compute the exchange amplitude we sum over all three-point interactions connected with the correspondent propagator. }
\label{figure:exsum}
\end{figure}

We represented this process in figure \ref{figure:exsum}. Before moving to the contact terms, let us pause here for a moment to discuss some details.

\subsubsection*{Renormalizable and non-renormalizable interactions}
\label{subsubsec:renor}

In tables \ref{MMM2}-\ref{M2M2M0}, and in the vertices associated, we treated normalizable and non-renormalizable interactions in the same fashion, implicitly assuming that both contributions can be of the same order. This may look a bit unnatural from a Lagrangian perspective. When dealing with EFTs and higher dimensional operators in $d=4$ we  have
\begin{align}
\mathcal{L}_{\text{EFT}}=\sum\limits_ia_i\Lambda_i^{4-\text{dim}(\mathcal L^i)}\mathcal{L}^i_{d\leq4}+\sum\limits_j\frac{b_j}{\Lambda_j^{\text{dim}\left(\mathcal{L}^j\right)-4}}\mathcal{L}^j_{d>4}\, ,
\end{align}
with $a^i$ and $b^j$  constants, $\mathcal{L}^i_{d\leq4}$ are the renormalizable and $\mathcal{L}^j_{d>4}$  the non-renormalizable terms. Usually, it is natural to expect something like
\begin{align}
\label{eq:reandnore}
\{a_i, b_j\}&\sim \mathcal{O}\left(1\right)\, , 	&	\Lambda_i^{4-\text{dim}(\mathcal L^i)}\mathcal{L}^i_{d\leq4}\gg\frac{\mathcal{L}^j_{d>4}}{\Lambda_j^{\text{dim}\left(\mathcal{L}^j\right)-4}}\, ,
\end{align}
which makes quite unlikely that diagrams coming from non-renormalizable parts in the Lagrangian can cancel the ones coming from renormalizable operators, unless some fine tunning occurs. As a warm-up, when we present our results in section \ref{sec:results} we will begin by studying a scenario like this. 

Nevertheless, since we want to be as general as possible, we will also investigate the case in which some $b_j$ can be $b_j\gg 1$, potentially making the contribution of the non-renormalizable pieces in the Lagrangian of the same order of the  renormalizable interactions. This is discussed extensively in section \ref{subsec:general}

\subsubsection*{Fierz-Pauli coupling}
\label{subsubsec:fp}

An important role for us is played by the vertex produced when we (minimally) couple the Fierz-Pauli action to gravity. This vertex must be non-vanishing in any theory of a massive spin-2 particle. The Fierz-Pauli Lagrangian, which describes the propagation of a massive spin-2 particle, denoted $h_{\mu\nu}$, in a flat background $\eta_{\mu\nu}$, is
 \begin{align}
\mathcal{L}=-\partial_\mu h^{\mu\nu}\partial_\nu h+\partial_\mu h^{\rho\sigma}\partial_\rho h^\mu_\sigma-\frac{1}{2}\partial_\mu h^{\rho\sigma}\partial^{\mu}h_{\rho\sigma}+\frac{1}{2}\partial_\mu h\partial^\mu h-\frac{1}{2}m^2\left(h^{\mu\nu}h_{\mu\nu}-h^2\right)\, .\label{eq:fplin}
\end{align}
Here $h=\eta^{\mu\nu}h_{\mu\nu}$.
Let us now consider a general background with a metric $\tilde g_{\mu\nu}$. Lagrangian \eqref{eq:fplin} becomes
\begin{align}
\mathcal{L}=-\nabla_\mu h^{\mu\nu}\nabla_\nu h+\nabla_\mu h^{\rho\sigma}\nabla_\rho h^\mu _\sigma-\frac{1}{2}\nabla_\mu h^{\rho\sigma}\nabla^{\mu}h_{\rho\sigma}+\frac{1}{2}\nabla_\mu h\nabla^\mu h-\frac{1}{2}m^2\left(h^{\mu\nu}h_{\mu\nu}-h^2\right)\label{eq:fpcur}\, ,
\end{align}
where the indices are now raised and lowered with the  metric $\tilde g_{\mu\nu}$ and the derivatives have been replaced by covariant derivatives.\footnote{One can also include a term $\mathcal{L}\supset\frac{R}{4}\left(h^{\mu\nu} h_{\mu\nu}-\frac{1}{2}h^2 \right)$ if the background has non-vanishing curvature $R$. It is easy to check that this term does not contribute to the massive-2, massive-2, massless-2 three-point interactions.} To obtain the three-point interactions between two massive and one massless spin-2 particles we expand the metric as
\begin{align}
\tilde g_{\mu\nu}&=\eta_{\mu\nu}+\frac{g_{\mu\nu}}{M_p}\, ,	&	\tilde g^{\mu\nu}&=\eta^{\mu\nu}-\frac{g^{\mu\nu}}{M_p}\, ,
\end{align}
and look for the terms of the form $hhg$. There are  three different on-shell contributions:
\begin{align}
h^{\mu\nu}h_{\mu\nu}=\left(\eta^{\alpha\mu}- \frac{g^{\alpha\mu}}{M_p}\right) \left(\eta^{\beta\nu}- \frac{g^{\beta\nu}}{M_p}\right) h_{\alpha\beta}h_{\mu\nu}|_{hhg}=-\frac{2}{M_p} g^{\alpha\mu}\eta^{\beta\nu} h_{\alpha\nu}h_{\mu\beta}=-2\frac{1}{M_p} B_{12}B_{13}B_{23}\, ,
\end{align}
also\footnote{See Appendix \ref{ap:fpver} for the details}
\begin{equation}
\label{yo}
\nabla_\mu h^{\rho\sigma}\nabla^{\mu}h_{\rho\sigma}|_{hhg}=\tilde g^{\mu\nu}\tilde g^{\rho\alpha}\tilde g^{\sigma\beta}\nabla_\mu h_{\rho\sigma}\nabla_\nu h_{\alpha\beta}|_{hhg}=-\mathcal{G}_1+2\frac{m^2}{M_p} B_{13}B_{23}B_{12}-2\mathcal{G}_2\, ,
\end{equation}
and in a similar way
\begin{equation}
\label{ye}
\nabla_\mu h^{\rho\sigma}\nabla_\rho h^\mu _\sigma|_{hhg}=\tilde g^{\mu\alpha}\tilde g^{\rho\nu}\tilde g^{\sigma\beta}\nabla_\mu h_{\rho\sigma}\nabla_\nu h_{\alpha\beta}|_{hhg}=\frac{1}{2}\mathcal{G}_3\, ,
\end{equation}
with $\mathcal{G}_3$ and $\mathcal{G}_2$ introduced in section \ref{sec:masimasimass}. This means that the Fierz-Pauli contribution to the on-shell three-point vertices can be written as:
\begin{equation}
\label{eq:universal}
\mathcal{G}_{FP}|_{hhg}\equiv \frac{1}{2}\left(\mathcal{G}_3+\mathcal{G}_1+2\mathcal{G}_2\right)\, ,
\end{equation}
which translates into 
\begin{align}
\label{eq:fpcons}
2g_1=g_2=2g_3\neq0\, .
\end{align}
This non-vanishing combination of three-point couplings must be satisfied in any theory of a massive spin-2 particle coupled to gravity. So in looking for possible consistent theories with appropriate Regge behaviour we are allowed to set any couplings to zero, but must demand that $g_1$, $g_2$ and $g_3$ are non-vanishing and satisfy the constraints (\ref{eq:fpcons}). Note that this matches the constraint from gauge invariance (\ref{eq:gaugeinva}).

\subsubsection*{Parity symmetry}
\label{subsubs:pari}

In the previous sections we introduced  parity-even and parity-odd three-point functions,  assuming that the theory can be parity violating. Given a particular choice of external polarisations, contributions from even-even and odd-odd vertices will be parity-even, whereas even-odd and odd-even connections will give rise to  parity-odd terms.

\begin{figure}[h]
\centering
\includegraphics[width=10cm]{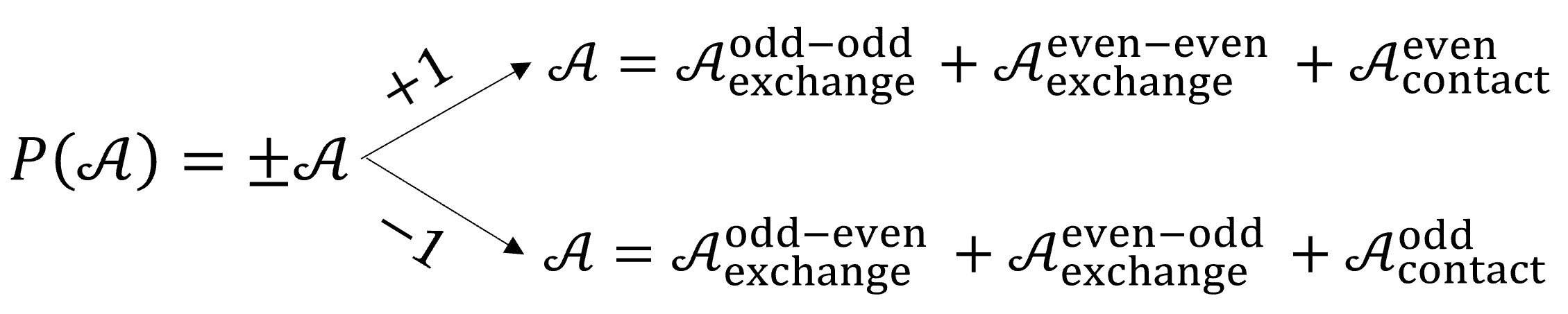}
\caption{In the transversity basis contributions with different parity decouple. }
\label{figure:par}
\end{figure}

One can  (and we will) avoid this mixing by using the transversity basis for the polarisations, see appendix \ref{ap:kin}, in which the spin of the particles is projected in the transverse direction to the scattering plane.\footnote{Unlike the usual helicity basis, in which spin and momentum are projected onto the same plane} In the transversity basis, the amplitudes $\mathcal A$ have definite parity $P$ and they transform under this symmetry as \cite{deRham:2017zjm}
\begin{align}
\label{eq:par}
P: \mathcal A_{\tau_1 \tau_2, \tau_3 \tau_4}\rightarrow (-1)^{\tau_1+\tau_2-\tau_3-\tau_4}\mathcal A_{\tau_1 \tau_2, \tau_3 \tau_4}\, ,
\end{align}
where the indices $_{\tau_1\tau_2,\tau_3\tau_4}$ refer to the helicities $\tau_i=\{0\, ,\, \pm1\, ,\, \pm2\}$ of the ingoing, 1 and 2, and outgoing, 3 and 4, particles. Regarding the exchange diagrams, this means that  when the sum of the helicities of the scattered particles is an even number, only even-even and odd-odd terms  contribute; on the contrary, if this sum is an odd number, one must consider even-odd and odd-even interactions. The same reasoning applies to the contact terms: in the transversity basis they \emph{decouple} according to the parity, as represented in figure \ref{figure:par}.
For our proposes and to prove our result it will be enough just to study the parity-even amplitudes.

Let us also comment that we exploit other nice properties of the transversity basis, like the simple form of the crossing symmetry relations. A review of transversity amplitudes and their properties is given in  \cite{deRham:2017zjm, osti_4128791,Bonifacio:2018vzv }.

\subsection{Contact terms}

\label{sec:contactterms}

Computing all the contact terms is the  trickiest and most cumbersome part since, in principle, there is an infinite number of them. To have a situation we can handle, we will only include interactions with an \emph{arbitrary but finite} number of derivatives, following the algorithm developed in  \cite{Bonifacio:2018vzv, Bonifacio:2018aon, Bonifacio:2019mgk}. As discussed in the introduction, neglecting the possibility of an infinite number of correlated higher derivative operators is part of the setting of an isolated massive spin-2 particle that we are studying. 

As explained above, we will only need to consider parity-even contact interactions. Parity-even and parity-odd contributions decouple in the transversity basis:  when the sum of the helicities of the scattered particles is an even (odd) number only parity-even (parity-odd) terms contribute. To reach our results it is enough to look at the parity-even amplitudes.\footnote{The reader interested in the parity-odd 4-point interactions is referred to \cite{Bonifacio:2018vzv}.}

We can start with the renormalizable (parity-even) operators. There are only two such operators:
\begin{subequations}
\label{eq:renorcontact}
\begin{align}
\label{eq:renorcontact1}
&B_{13}B_{14}B_{23}B_{24}+B_{12}B_{14}B_{23}B_{34}+ B_{12}B_{13}B_{24}B_{34}\, , \\\label{eq:renorcontact2} 
&B_{14}^2B_{23}^2+B_{12}^2B_{34}^2+B_{13}^2B_{24}^2\, ,
\end{align}
\end{subequations}
or, written in the Lagrangian basis,
\begin{subequations}
\label{eq:renorcontactl}
\begin{align}
\label{eq:renorcontact1l}
&h^{1\mu\nu}h^{3\alpha}_\nu h^{2\beta}_\alpha h^{4}_{\beta\mu}+h^{1\mu\nu}h^{2\alpha}_\nu h^{3\beta}_\alpha h^{4}_{\beta\mu}+h^{1\mu\nu}h^{2\alpha}_\nu h^{4\beta}_\alpha h^{3}_{\beta\mu}\, , \\
\label{eq:renorcontact2l} 
&h^{1\mu\nu} h^4_{\mu\nu}\, h^{2\alpha\beta} h^3_{\alpha\beta}+h^{1\mu\nu} h^2_{\mu\nu}\, h^{3\alpha\beta} h^4_{\alpha\beta}+h^{1\mu\nu} h^3_{\mu\nu}\, h^{2\alpha\beta} h^4_{\alpha\beta}\, .
\end{align}
\end{subequations}

Non-renormalizable contact terms, on the other hand, are many. Schematically, the list of parity-even contact terms with up to $N$ derivatives  can be written by first computing   
\begin{align}
\label{eq:tensorstruc}
V_{2,2,2,2}\left(m,m,m,m\right)=\sum\limits_{n,m,k,l}^{n+k+m+l=N}\sum\limits_{\text{ contractions}}\frac{a_{nmkl}}{\Lambda^{ n+m+k+l}}\partial^n h\,  \partial^m h\, \partial^k h\, \partial^l h\, ,
\end{align}
with the first summation taking into account all the possible ways in which the indices of the four $h_{\mu\nu}$ and  the derivatives $\partial_\mu$ can be contracted, and then symmetrizing under the interchange of any two particles. We are using  in \eqref{eq:tensorstruc} the description in terms of fields, instead of $\{A_{ij}, B_{mn}\}$, to make the analysis more understandable.

The above discussion serves to illustrate the complexity of the problem. In practice, though, we will follow a slightly different approach, adapting the strategy developed in \cite{Bonifacio:2018vzv}. In short, one \emph{only} needs to compute explicitly the tensor structures\footnote{By tensor structures in this context we mean the terms $\sum\limits_{n,m,k,l} \partial^n h\,  \partial^m h\, \partial^k h\, \partial^l h$ with the indices of the derivatives always  contracted with an index of some $h$. } invariant under the group of permutations that preserve the Mandelstam variables invariant, the so-called kinetic permutations $\Pi^{\text{kin}}$ \cite{Kravchuk:2016qvl}
\begin{align}
\label{eq:permutationsym}
\Pi^{\text{kin}}=\{\mathds{1},\, (12)(34), \, (13)(24), \, (14)(23)\}\, ,
\end{align}
where $(ij)(kl)$ means that we do the permutations $i\leftrightarrow j$ and $k\leftrightarrow l$ at the same time. There are 201 such (parity-even) tensor structures, which we will denote  by $\mathds{T}_a$ and that we list in appendix \ref{ap:contactterms}.\footnote{Only 97 of them are independent in $d=4$, since the vanishing of the Grassmannian of the vectors $\{p_i,\epsilon_{i\mu}\}$ imposes extra constraints. However, it is harder to find a basis of the independent structures than to work with the redundancies.} Then, each of these structures is multiplied by an arbitrary polynomial $f_{a}\left(s,t\right)$ of $s$ and $t$, yielding:
\begin{align}
\label{eq:contacttermsexplici}
\boxed{\tilde A^{\text{contact}}=\sum\limits_a^{201} f_{a}\left(s,t\right)\mathds{T}_a}\, .
\end{align}
We are calling this intermediate result $\tilde A^{\text{contact}}$ to distinguish it from the actual $ A^{\text{contact}}$, invariant under the permutation of any two particles. Indeed, to obtain $ A^{\text{contact}}$, one has to impose the remaining permutation symmetries in $\tilde A^{\text{contact}}$. These other permutations, under which $(s,t,u)$ are not invariant, will lead to crossing relations that must be satisfied. In section \ref{subsec:general} we will explain extensively what are these crossing constraints and how to obtain $ A^{\text{contact}}$.


\section{Constraining theories through Regge growth}
\label{sec:results}

We consider the most general effective theory of a massive spin-2 particle and impose the constraint (\ref{crgc}). The result we are after is whether such a theory, satisfying (\ref{crgc}), exists. At least, it must contain the couplings (\ref{eq:universal}) as non-vanishing, and so satisfy (\ref{eq:fpcons}). That is the starting point.  The rest of the couplings are allowed to take any values, including vanishing. The analysis follows a simple procedure:
\begin{enumerate}
\item Compute all exchange and contact contributions for all polarisations. This results in some function of $s$ and $t$
\begin{align}
\mathcal{A}_{2\rightarrow 2}=\sum\limits_i\mathcal{A}_i^{\text{exchange}}+\sum\limits_j\mathcal{A}_j^{\text{contact}}\equiv\mathcal{A}\left(s,t\right)\, .
\end{align}

\item Take the limit $\{s\gg |t|,m^2\}$. Expand $\mathcal{A}\left(s,t\right)$ in powers of $s$
\begin{align}
\lim_{s\gg t,m^2} \mathcal{A}\left(s,t\right)=\dots + \mathcal{A}_2\left(t\right)s^2+ \mathcal{A}_3\left(t\right)s^3+\dots \mathcal{A}_n\left(t\right)s^n\, .
\end{align}
\item Impose 
\begin{align}
\label{eq:crgcondition}
\mathcal{A}_m\left(t\right)&=0\, , \,\, \text{for }\,\,  m\geq 3\, ,\,\, \forall\, t<\, 0\, ,
\end{align}
and for any external polarisation configuration. Equation \eqref{eq:crgcondition} will yield several sub-equations, one for each power of $t^i$. These sub-equations will depend linearly on the coefficients of the contact terms  and quadratically on the three-point couplings.
\item Solve the previous equations. If (\ref{crgc}) can be satisfied non-trivially, some relation between the three-point and the four-point couplings will be obtained. Otherwise, no possible theory satisfying (\ref{crgc}) exists. 
\end{enumerate}

\subsection{A maximally natural scenario}
\label{subsec:naturalscenario}

As a warm-up, we will start by considering a simplified scenario in which all dimensionless couplings are taken to be of order one. So this means that the magnitude of operators is controlled by their dimension and the cutoff scale $\Lambda$. So this is a type of maximally natural scenario. 

The starting point is the demand that the combination of terms in \eqref{eq:universal} is non-vanishing. These operators have all mass dimension five.  By direct computation, it is easy to check that with only these three pieces, the four-point scattering of a massive spin-2 particle can  never satisfy (\ref{crgc}).

We therefore need to see if the other operators can be chosen to cancel the too-fast Regge growth. The ``maximally natural'' scenario we are considering means that such a possible cancellation is very restricted.
Regarding the contact terms, only the ones with mass dimension less than or equal to six can be useful. Higher dimensional operators will be suppressed by higher powers of $\Lambda$, which makes them parametrically smaller. We would have equations like 
\begin{align}
a \frac{s^3}{\Lambda^2 m^4}+b\frac{ s^3}{\Lambda^4m^2}+c\frac{ s^3}{\Lambda^6}+\dots =0\, ,
\end{align}
which,  unless  the coupling constants satisfy $c\gg b \gg a$,  as we are forbidding in this section, must be solved order by order.\footnote{$\{a, b, c\}$ serve as an illustrative example, we are not referring to any couplings in particular.}
For the same reasons, we can ignore three-point vertices suppressed by  $\Lambda^n$ for $n\geq 2$. 

This leaves us with the following three-point vertices:
\begin{subequations}
\begin{align}
&V_{2,2,2}\left(m,m,m\right)=a_1 \mathcal{M}_1+a_2 \mathcal{M}_2+a_3 \mathcal{M}_3+a_5 \mathcal{M}_5\, ,	 \\	&V_{2,2,2}\left(m,m,0\right)=g_1 \mathcal{G}_1+g_2 \mathcal{G}_2+g_3 \mathcal{G}_3+g_6 \mathcal{G}_6+g_7 \mathcal{G}_7\, , \\	 	
&V_{2,2,1}\left(m,m,m_1\right)=k_1 \mathcal{K}_1+k_3 \mathcal{K}_3\, ,\\ 	
&V_{2,2,0}\left(m,m,M\right)=s_1\mathcal{S}_1+s_2\mathcal{S}_2+s_4\mathcal{S}_4\, .
\end{align}
\end{subequations}
Four-point interactions with up to two derivatives have to be included. Focusing only on the parity-even terms, there are 12 distinct choices:
\begin{itemize}
\item 2 contact terms with no derivatives, already discussed in  \eqref{eq:renorcontact}. We multiply them by $\{c_1,\, c_2\}$ .
\item 4 contact terms with two derivatives with the indices of the derivatives contracted among themselves (terms of the form $\partial^\mu h\partial_\mu h h h\, +\,  \text{symmetrization}$). They come with the couplings $\{c_3,\, \dots\, , c_6\}$ .
\item 6 contact terms with two derivatives with the indices of the derivatives contracted with the massive spin-2 fields (terms of the form $\partial^\mu\partial^\nu h_{\mu\nu} h h h\, +\,  \text{symmetrization}$). We use $\{c_7,\, \dots\, , c_{12}\}$  to parameterize these interactions.
\end{itemize}
Overall, there are $12+14$ degrees of freedom. We will show now that this system can never satisfy (\ref{crgc}), writing some intermediate steps.  We will indicate by $ij\rightarrow ab$ the helicity of the  particles scattered, so taking values $\{ 0,\, \pm1,\, \pm2\}$.
\begin{itemize}
\item From $11\rightarrow 11$  at  order $s^4 t$:
\begin{align}
\frac{a_2^2}{m^8M_p^2}=0\,\, \longrightarrow \,\, a_2=0\, .
\end{align}
\item From $00\rightarrow 00$    at order  $s^6 t$ (using $a_2=0$):
\begin{align}
\frac{a_3^2}{m^{12}M_p^2}=0\,\, \longrightarrow \,\, a_3=0\, .
\end{align}
\item From $(-1)(2)\rightarrow (-1)(-2)$ at order $s^3\sqrt{t}$ (using $a_2=0=a_3$):
\begin{align}
\frac{a_5^2+3 \left(4 (g_2-2 g_3)^2+(g_7-2 g_6)^2\right)}{m^5 \text{Mp}^2}=0\,\, \longrightarrow \,\, a_5=0\, , \, g_2=2g_3\, , \, g_7=2g_6
\end{align}
\item From the equations involving just contact couplings
\begin{align}
\{c_7,\, \dots\, , c_{12}\}=0\, .
\end{align}
\item Plugging  these constraints in the other equations the only real solution is:
\begin{align}
\{a_1,a_5,g_1,g_3,g_5,g_6,s_2,k_1,k_3\}=0\,,
\end{align}
with $s_1$ being a free parameter. 
\end{itemize}
This means that all cubic couplings, except for the renormalizable interaction with the scalar field, must vanish. In particular, the required combination (\ref{eq:universal}) also, and so the theory is trivial. In other words, under the assumptions made, there cannot exist a theory of a massive spin-2 particle consistent with (\ref{crgc}).

\subsection{The general case}   
\label{subsec:general}

After discussing a simplified version of the problem, we will now tackle  the general case, in which all exchange  and  contact diagrams with any finite number of derivatives are included with arbitrary coefficients. Our strategy  follows closely, though not identically, the one developed in \cite{Bonifacio:2018vzv, Bonifacio:2018aon, Bonifacio:2019mgk}. 

In what follows, we will describe in detail the procedure without showing any explicit computation. The reader interested can find an ancillary \texttt{Mathematica} file attached to the paper with the code used. We only focus on the parity-even choices of polarisations, recall equation \eqref{eq:par}, since this is enough to arrive at our results.

To start with, let us split the four-point scattering amplitude $ A_{\tau_1\tau_2,\tau_3\tau_4}$ into the contact and the exchange contributions:
\begin{align}
    A_{\tau_1\tau_2,\tau_3\tau_4}=A^{\text{exchange}}_{\tau_1\tau_2,\tau_3\tau_4}+A^{\text{contact}}_{\tau_1\tau_2,\tau_3\tau_4}\, ,
\end{align}
where again the indices $_{\tau_1\tau_2,\tau_3\tau_4}$ refer to the helicities $\tau_i=\{0\, ,\, \pm1\, ,\, \pm2\}$ of the ingoing, 1 and 2, and outgoing, 3 and 4, particles. Then, the steps to follow are:
\begin{itemize}
    \item[1.]  Compute $A^{\text{exchange}}_{\tau_1\tau_2,\tau_3\tau_4}$. As commented in in section \ref{sec:threepoint} one needs to take two sets of the three-point vertices, 
    \begin{subequations}
    \begin{align}
&V_{2,2,2}\left(m,m,m\right)=a_1 \mathcal{M}_1+a_2 \mathcal{M}_2+a_3 \mathcal{M}_3+a_4\mathcal{M}_4+a_5\mathcal{M}_5+a_6\mathcal{M}_6\, ,\\	 	
&V_{2,2,2}\left(m,m,0\right)=g_1 \mathcal{G}_1+g_2 \mathcal{G}_2+g_3 \mathcal{G}_3+g_{4} \mathcal{G}_{4}+g_5 \mathcal{G}_5+g_6 \mathcal{G}_6+g_7 \mathcal{G}_7\nonumber\\&+g_8 \mathcal{G}_8+g_9 \mathcal{G}_9\, ,	\\	&V_{2,2,1}\left(m,m,m_1\right)=k_1 \mathcal{K}_1+k_2 \mathcal{K}_2+k_3 \mathcal{K}_3+k_4 \mathcal{K}_4\, ,\\	
&V_{2,2,0}\left(m,m,M\right)=s_1\mathcal{S}_1+s_2\mathcal{S}_2+s_3\mathcal{S}_3+s_4\mathcal{S}_4+s_5\mathcal{S}_5\, ,
\end{align}
    \end{subequations}
``remove" the  exchanged particle and connect them through the correspondent propagator. We listed the functions and conventions we used in appendix \ref{ap:kin}. The result of this step is an amplitude that depends on the product of the twenty-four three-point coupling constants.
    \item[2.]Take an ansatz for $A^{\text{contact}}_{\tau_1\tau_2,\tau_3\tau_4}$ with the kinematical singularities  \cite{osti_4128791,Cohen-Tannoudji:1968lnm, kota} factored out
    \begin{align}
    \label{eq:contansa}
A^{\text{contact}}_{\tau_1\tau_2,\tau_3\tau_4}\left(s,t\right)=\frac{a^{\text{contact}}_{\tau_1\tau_2,\tau_3\tau_4}\left(s,t\right)+i\sqrt{stu}\, b^{\text{contact}}_{\tau_1\tau_2,\tau_3\tau_4}\left(s,t\right)}{\left(s-4m^2\right)^{|\sum\limits_i\tau_i|/2}}\, ,
    \end{align}
with $a^{\text{contact}}_{\tau_1\tau_2,\tau_3\tau_4}=\sum\limits_{n,m}\, \alpha^{\tau_1\tau_2,\tau_3\tau_4}_{m,n}\, t^m\,  s^n$ and $b^{\text{contact}}_{\tau_1\tau_2,\tau_3\tau_4}=\sum\limits_{i,j}\, \beta^{\tau_1\tau_2,\tau_3\tau_4}_{i,j}\, t^i\, s^j$  arbitrary polynomials   at this point. The indices $\{ n,m\}$ and $\{ i,k\}$ run from 0  to $n+m=p$ and $i+j=k$, with $p$ and $k$ such that in the Regge limit, $\{s\gg |t|\,\}$, the real and imaginary parts of both $A^{\text{exchange}}$ and $A^{\text{contact}}$ have the same energy scaling. The particular form of the ansatz follows from the fact that we are using the transversity basis for the polarisations, a basis which we  introduced  in section \ref{subsubs:pari}.
\item[3.] Constrain $a^{\text{contact}}_{\tau_1\tau_2,\tau_3\tau_4}$ and $b^{\text{contact}}_{\tau_1\tau_2,\tau_3\tau_4}$ by requiring that, in the Regge limit, the total amplitude $A_{\tau_1\tau_2,\tau_3\tau_4}$ does not grow faster than $s^2$. In practice this means:
\begin{align}
\label{eq:reglimitconstraint}
\lim_{s \to \infty\, ,\, \forall t<0} \left[\frac{A^{\text{exchange}}_{\tau_1\tau_2,\tau_3\tau_4}}{s^3}+\frac{A^{\text{contact}}_{\tau_1\tau_2,\tau_3\tau_4}}{s^3}\right]=0\, .
\end{align}
which gives a series of relations between the contact couplings $\{\alpha^{\tau_1\tau_2,\tau_3\tau_4}_{m,n},\, \beta^{\tau_1\tau_2,\tau_3\tau_4}_{m,n}\}$ and the product of cubic couplings $\{a_i,g_j,k_m,s_n\}$. By rewriting these products as new  variables
\be
a_i a_j\equiv a_{ij}\, ,\,g_i g_j\equiv g_{ij}\, ,\,k_i k_j\equiv k_{ij}\, ,\,s_i s_j\equiv s_{ij}\,,
\ee
one can linearise the equations resulting from \eqref{eq:reglimitconstraint}, making  the problem computationally more tractable.  The condition \eqref{eq:reglimitconstraint} is different from the one imposed in  \cite{Bonifacio:2018vzv, Bonifacio:2018aon, Bonifacio:2019mgk}, who were interested in the case $\lim_{\{s \to \infty\, ,\, -t\to \infty\}} \frac{A_{\tau_1\tau_2,\tau_3\tau_4}}{s^3}=0$ with $s/t$ fixed.

\item[4.] At this stage we need to impose further restrictions on the contact interactions which come from crossing symmetries.\footnote{Crossing symmetries are the permutation symmetries that do change the Mandelstan variables. Recall the discussion in section \ref{sec:contactterms}} Imposing  the crossing symmetries on the contact terms requires\cite{deRham:2017zjm,Kotanski:1965zz} 
\begin{subequations}
\label{eq:a}
\begin{align}
\mathcal{A}^{\rm contact}_{\tau_1 \tau_2, \tau_3 \tau_4}(s,t) & = e^{i\left(\pi- \chi_t\right) \sum_j \tau_j }\mathcal{A}^{\rm contact}_{-\tau_1 -\tau_3, -\tau_2 -\tau_4}(t,s),\\
\mathcal{A}^{\rm contact}_{\tau_1 \tau_2, \tau_3 \tau_4}(s,t) & = e^{i \left( \pi- \chi_u\right) \sum_j \tau_j }\mathcal{A}^{\rm contact}_{-\tau_1 -\tau_4, -\tau_3 -\tau_2}(u,t),
\end{align}
\end{subequations}
where
\begin{align}
e^{-i \chi_t} \equiv \frac{-st -2 i m \sqrt{s t u}}{\sqrt{ s(s-4m^2)t(t-4m^2)}}, \quad e^{-i \chi_u} \equiv \frac{-su +2 im \sqrt{s t u}}{\sqrt{ s(s-4m^2)u(u-4m^2)}}.
\end{align} 
These equations relate  some of the surviving $\{\alpha^{\tau_1\tau_2,\tau_3\tau_4}_{m,n},\, \beta^{\tau_1\tau_2,\tau_3\tau_4}_{m,n}\}$ to the constants  $\{c_{ij}, \, g_{ij},\, k_{ij},\, s_{ij}\}$. As explained some steps before, this elegant form of the crossing symmetries \eqref{eq:a} is a feature of the transversity basis.

\item[5.] Make the ansatz for $\mathcal{A}^{\rm contact}$ consistent with the  expression \eqref{eq:contacttermsexplici}. This should hold for all polarisations, so that 
\be
\mathcal{A}^{\rm contact}_{\tau_1 \tau_2, \tau_3 \tau_4}=\sum\limits_a^{201} f_{a}^{\tau_1 \tau_2, \tau_3 \tau_4}\left(s,t\right)\mathds{T}_a^{\tau_1 \tau_2, \tau_3 \tau_4} \;,
\ee
for all 
\be
\tau_1 +\tau_2- \tau_3 -\tau_4=k\, , \;\;\;\;\;\, k\in2\mathds{Z} \;.
\ee
To do this, omitting for clearness the polarisation indices, one needs to expand the polynomial $f_{a}\left(s,t\right)=\sum\limits_{ij}f_{ij}t^is^j$ and match it at each order with the result obtained in the previous step.\footnote{ Note that for $\tau_1 +\tau_2- \tau_3 -\tau_4=n\, , \, n\in\left(2\mathds{Z}+1\right)$ one should do the same computation but using the parity-odd tensor structures, which we are not using here.} Finally, one has to undo the change of variable in the product of cubic couplings and make sure that everything is consistent: there cannot be contradictions, e.g.  $c_{ii}=0,\, c_{ij}\neq0$, and all couplings must be real. After  this step, we end up with the most general $2\rightarrow2$ scattering of a massive spin-2 particle compatible with the Regge growth requirement (\ref{crgc}). 
\end{itemize}

After implementing this 5-step procedure, we find that the Regge growth bound (\ref{crgc}) can only be satisfied if:
\begin{align}
&a_i=0\,,\,  \forall i \, ; 	&	& g_j=0\,,\,  \forall j\geq 2\, ; 	&	& k_l=0\,,\,  \forall l\, ;  	&	& s_n=0 \,,\,  \forall n\geq 2\,  .
\end{align}
Combining this result with the fact that interactions with gravity must include the Fierz-Pauli action, see equation \eqref{eq:fpcons}, we arrive at \footnote{The same result follows by using  equation \eqref{eq:gaugeinva} instead of \eqref{eq:fpcons}, which holds in any theory in which the graviton self-interactions include those from the usual Einstein-Hilbert term.} 
\begin{align}
&\boxed{a_i=0\,,\,  \forall i }\,  	&	& \boxed{g_j=0\,,\,  \forall j}
  	&	& \boxed{k_l=0\,,\,  \forall l}\,   	&	& \boxed{s_n=0 \,,\,  \forall n\geq 2}\, \; .
\end{align}
So all Lagrangian terms must vanish (with the sole exception of the scalar coupling $h^{\mu\nu}h_{\mu\nu}\phi$). Hence, there is no such theory.
We therefore arrive at the main result of this paper: theories containing a single massive spin-2 particle are not compatible with (\ref{crgc}). Recall that this follows under the following assumptions:
\begin{itemize}
\item The massive spin-2 particle is allowed to couple to a massless spin-2, a massive spin-1 and a massive or massless scalar particle. 
\item  All contact terms with any finite number of derivatives are taken into account, but infinite series of contact terms which can re-sum to poles are not included.
\end{itemize}

\section{Summary}
\label{s:conclu}

As discussed in the introduction, it is simple to summarise our results: we find that there are no possible theories with a massive spin-2 particle coupled to gravity, and a gap to the next spin-2 (or higher) states, which satisfy the constraint (\ref{crgc}). So a theory of an isolated massive spin-2 particle, coupled to gravity, will necessarily exhibit Regge growth which is faster than $s^2$ in the gap between the massive spin-2 particle and any further spin-2 (or higher) states. 

This result holds up to the assumption of neglecting infinite series of higher derivative terms which can resum into a pole. As explained, this is justified because such a series would only be relevant near the mass scale of the spin-2 (or higher) particle which was integrated out to generate it, and we assume a gap to any such scales. 

Our results strongly suggest that theories of isolated massive spin-2 particles, coupled to gravity, are in the Swampland. To prove this one would need to prove that the fast Regge scaling (\ref{crgc}) is inconsistent. This is essentially what the Classical Regge Growth Conjecture proposes \cite{Chowdhury:2019kaq}. As well as what is expected from general Regge growth bounds, for example as studied in \cite{Haring:2022cyf}. 

Of course, theories with massive spin-2 states do occur in ultraviolet complete consistent theories, but they are always part of an infinite tower. In restricted settings, for example when pure gravity is dimensionally reduced, it is possible to bound the gap between massive spin-2 states \cite{Bonifacio:2019ioc}. We expect that such a bound can be deduced even for the most general possible theories, as studied in this paper. We leave a proof of this for future work. 

Finally, let us note that theories where there is no massless spin-2 state, but only an isolated massive spin-2 one, so theories of massive gravity, also exhibit the fast Regge growth we have found. This follows from our results, which are that the only couplings (so non-quadratic) interactions which are compatible with slow Regge growth are the coupling to a scalar field.

\bigskip

\centerline{\bf  Acknowledgments}

\vspace*{.5cm}

We would like to thank  James Bonifacio and Kurt Hinterbichler for very helpful discussions and explanations and  Shiraz Minwalla for useful discussions. The work of EP and JQ is supported by the Israel Science Foundation (grant No. 741/20). The work of EP is supported by the Israel planning and budgeting committee grant for supporting theoretical high energy physics. The work of SK, EP and JQ is supported by the German Research Foundation through a German-Israeli Project Cooperation (DIP) grant ``Holography and the Swampland". The work of SK is also supported in part by an ISF, centre for excellence grant (grant number 2289/18), Simons Foundation grant 994296, and Koshland Fellowship.


\appendix

\section{Kinematics and convention}
\label{ap:kin}

In this appendix we will  write explicitly the definitions of the kinematical variables used to compute the $2\rightarrow 2$ scattering of the massive spin-2 particle. We will use the conventions of  \cite{Bonifacio:2018vzv, Bonifacio:2018aon, Bonifacio:2019mgk}.

The incoming particles are labelled by $1$ and $2$, and outgoing particles by $3$ and $4$. Momentum conservation requires
\begin{align}
p_1+p_2=p_3+p_4\, ,
\end{align}
where we take
\begin{align}
p_i^\mu=\left(E,p \sin\theta_j,0,p\cos\theta_j\right)\, ,
\end{align}
with $E^2=p^2+m^2$ and $\theta_1=0$, $\theta_2=\pi$, $\theta_3=\theta$, $\theta_4=\theta-\pi$. The Mandelstam variables are
\begin{align}
\label{mandelstam}
s=&-\left(p_1+p_2\right)^2\, ,	&	t&=-\left(p_1-p_3\right)^2\, ,	&	u&=-\left(p_1-p_4\right)^2\, ,
\end{align}
and we are taking the metric $\eta=\text{diag}\left(-1,1,1,1\right)$. They satisfy
\begin{align}
s+t+u=4m^2\, ,
\end{align}
with $m$ the mass of the spin-2 particle, and are related to $E$ and $\theta$ by
\begin{align}
s&=4E^2\, ,	&	cos\theta=1-\frac{2t}{4m^2-s}\, .
\end{align}
For the three polarisation vectors of the massive spin-1 particles we use the so-called transversity basis \cite{deRham:2017zjm, Kotanski:1965zz, osti_4128791} in which the spin of the particles are projected on the axis orthogonal to the scattering plane
\begin{subequations}
    \begin{align}
\epsilon^{\mu}_{\left(\pm1\right)}\left(p_j\right)&=\frac{i}{\sqrt{2}m}\left(p,E\sin\theta_j\pm im \cos\theta_j,0,E\cos\theta_j\mp im\sin\theta_j\right)\, ,\\
\epsilon^{\mu}_{\left(0\right)}\left(p_j\right)&=\frac{1}{m}\left(0,0,1,0\right)\, .
\end{align}
\end{subequations}
They are transverse and orthonormal. From them one can construct the five polarisation tensors of massive spin-2 particles
\begin{subequations}
\label{eq:pol2def}
    \begin{align}
\epsilon^{\mu\nu}_{\left(\pm2\right)}&=\epsilon^{\mu}_{\left(\pm1\right)}\epsilon^{\nu}_{\left(\pm1\right)}\, ,\\
\epsilon^{\mu\nu}_{\left(\pm1\right)}&=\frac{1}{\sqrt{2}}\left(\epsilon^{\mu}_{\left(\pm1\right)}\epsilon^{\nu}_{\left(0\right)}+\epsilon^{\mu}_{\left(0\right)}\epsilon^{\nu}_{\left(\pm1\right)}\right)\, ,\\
\epsilon^{\mu\nu}_{\left(0\right)}&=\frac{1}{\sqrt{6}}\left(\epsilon^{\mu}_{\left(1\right)}\epsilon^{\nu}_{\left(-1\right)}+\epsilon^{\mu}_{\left(-1\right)}\epsilon^{\nu}_{\left(1\right)}+2\epsilon^{\mu}_{\left(0\right)}\epsilon^{\nu}_{\left(0\right)}\right)\, ,\\
\end{align}
\end{subequations}
which are transverse, traceless and orthonormal.

Regarding the propagators, the propagator of a scalar particle with mass $M$ is
\begin{align}
    \frac{-i}{p^2+M^2-i\epsilon}\, .
\end{align}
For a massive spin-1 particle, first we need to introduce the projector
\begin{align}
    \Pi_{\mu\nu}\left(\tilde m\right)=\eta_{\mu\nu}+\frac{p_\mu p_\nu}{\tilde m^2}\, ,
\end{align}
from which one can write the propagator of a  massive spin-1 particle with mass $m_1$ as
\begin{align}
  P_{\mu\nu}= \frac{ -i\Pi_{\mu\nu}\left( m_1\right)}{p^2+m_1^2-i\epsilon}\, .
\end{align}
Finally, the propagator of a massive spin-2 particle of mass $m$ is
\begin{align}
     P_{\mu_1\mu_2,\nu_1\nu_2}=\frac{-i}{2}  \frac{\Pi_{\mu_1\nu_1}\Pi_{\mu_2\nu_2}+\Pi_{\mu_1\nu_2}\Pi_{\mu_2\nu_1}-\frac{2}{3}\Pi_{\mu_1\mu_2}\Pi_{\nu_1\nu_2}}{p^2+m^2-i\epsilon}\, ,
\end{align}
with $\Pi_{\nu_1\nu_2}=\Pi_{\nu_1\nu_2}\left( m\right)$, whereas for a massless spin-2 (in de Donder gauge) it reads 
\begin{align}
     \tilde P_{\mu_1\mu_2,\nu_1\nu_2}=\frac{-i}{2}  \frac{\eta_{\mu_1\nu_1}\eta_{\mu_2\nu_2}+\eta_{\mu_1\nu_2}\eta_{\mu_2\nu_1}-\eta_{\mu_1\mu_2}\eta_{\nu_1\nu_2}}{p^2-i\epsilon}\, .
\end{align}

\section{On-shell quantities}
\label{ap:onshell}

In the first part of this appendix we explain how to construct the on-shell three-point vertices presented in section \ref{sec:threepoint}, following the procedure developed in \cite{Costa:2011mg}. In the second part of the appendix \ref{ap:pd3pfunc} we list all possible vertices\footnote{For the coupling massive-2-, massive-2, scalar, the table \ref{M2M2M0} already contained all terms, so we will not discuss it here.}.

\subsection{General discussion}

Let us start by fixing the notation. We are interested in the on-shell three-point interactions of 3 particles with masses $m_i$, momenta $p_i$, spin $s_i$ and polarisation tensors $\epsilon_i$, with $i=1,2,3$ respectively. They satisfy $\epsilon_i\cdot p_i =0$, whereas the fact that they are on shell means $p_i^2=p_i\cdot p_i=-m_i^2$. Our convention for momentum conservation reads $\sum\limits_i p_i =0$. As explained at the beginning of section \ref{sec:ampli} we will formally write the polarisation matrices as the product of vectors $\epsilon_{i\mu\nu}=\epsilon_{i\mu}\epsilon_{i\nu}$, as a trick to keep track of the contractions more easily.

\subsubsection*{Parity even}

The parity-even on-shell three-point vertices are polynomial functions of the product of $\{\epsilon_{i\mu},p_j\}$, homogeneous  of order $s_a$  in each $\epsilon_{i\mu}$ -from now on, and making abuse of notation, we will call for simplicity $\epsilon_{i\mu}\equiv \epsilon_{i}$-. Taking into account that $\epsilon_i\cdot\epsilon_i=0=\epsilon_i\cdot p_i$ these interactions can be constructed  from nine  contractions  
\begin{align}
\epsilon_1 &\cdot \epsilon_2\, , & \epsilon_1 &\cdot \epsilon_3\, , & \epsilon_2 &\cdot \epsilon_3\, ,\nonumber \\
\epsilon_1 &\cdot p_2\, , & \epsilon_1 &\cdot p_3\, , & \epsilon_2 &\cdot p_1\, , \\
\epsilon_2 &\cdot p_3\, , & \epsilon_3 &\cdot p_1\, , & \epsilon_3 &\cdot p_2\nonumber\, ,
\end{align}
which, using $p_1+p_2+p_3=0$, can be reduced further to
\begin{align}
&\epsilon_1 \cdot \epsilon_2\, , & &\epsilon_1 \cdot \epsilon_3\, , & &\epsilon_2 \cdot \epsilon_3\, ,\nonumber \\
&\epsilon_1 \cdot p_2\, , & &\epsilon_3 \cdot p_1\, , &  &\epsilon_2\cdot p_1\, , \\
 &\cancelto{-\epsilon_1 \cdot p_2}{\epsilon_1 \cdot p_3}\, , & &\cancelto{-\epsilon_3 \cdot p_1}{\epsilon_3 \cdot p_2}\, ,	&	&\cancelto{-\epsilon_2 \cdot p_1}{\epsilon_2 \cdot p_3} \nonumber\, ,
\end{align}
and so the most general parity-even three-point interaction must take the form
\begin{align}
\label{eq:amplieven}
\mathcal{M}_{s_1,s_2,s_3}\left(p_1,p_2,p_3\right)=C\left(\epsilon_1\cdot\epsilon_2\right)^{x_{12}}\left(\epsilon_1\cdot\epsilon_3\right)^{x_{13}}\left(\epsilon_2\cdot\epsilon_3\right)^{x_{23}}\left(\epsilon_1\cdot p_2\right)^{y_{12}}\left(\epsilon_3\cdot p_1\right)^{y_{31}}\left(\epsilon_2\cdot p_1\right)^{y_{21}}\, ,
\end{align}
where $C$ is an arbitrary constant\footnote{Any product $p_i\cdot p_j$ can be rewritten as a function of the masses and reabsorbed in this constant}  and the exponents are non-negative integers that have to satisfy
\begin{subequations}
\label{eq:onconseven}
\begin{align}
x_{12}+x_{13}+y_{12}&=s_1\, , \\
x_{12}+x_{23}+y_{21}&=s_2\, , \\
x_{13}+x_{23}+y_{31}&=s_3\, , \\
s_i\geq 0\, .
\end{align}
\end{subequations}
In the particular case $d=4$, we can also use the fact that the Gram determinant of the vectors $\{\epsilon_1,\epsilon_2,\epsilon_3,p_1,p_2\}$ must vanish: five vectors cannot be all independent in four dimensions. We already exploded this fact in the main text. 

\subsubsection*{Parity odd}

On the other hand, parity-odd three-point functions are constructed from  contractions with the Levi-Civita tensor $\varepsilon$, $\varepsilon\left(p_i,p_j,\epsilon_k,\epsilon_l\right)\equiv \varepsilon_{\mu\nu\alpha\beta}\,p^{i\mu}\, p^{j\nu} \,\epsilon^{k\alpha}\,\epsilon^{l\beta}\,$ or $\varepsilon\left(p_i,\epsilon_j,\epsilon_k,\epsilon_l\right)\equiv \varepsilon_{\mu\nu\alpha\beta}\,\epsilon^{i\mu}\, \epsilon^{j\nu} \,\epsilon^{k\alpha}\,\epsilon^{l\beta}$,   plus some extra parity-even structures to include enough number of $\epsilon_i$. The most general parity-odd on-shell three-point amplitude can always be written as:
\begin{align}
\label{eq:ampliodd}
&\mathcal{M}_{s_1,s_2,s_3}\left(p_1,p_2,p_3\right)=\nonumber\\&D \varepsilon\left(p_1^{z_1},p_2^{z_2},\epsilon_1^{z_3},\epsilon_2^{z_4},\epsilon_3^{z_5}	,\right)\left(\epsilon_1\cdot\epsilon_2\right)^{\tilde x_{12}}\left(\epsilon_1\cdot\epsilon_3\right)^{\tilde  x_{13}}\left(\epsilon_2\cdot\epsilon_3\right)^{\tilde  x_{23}}\left(\epsilon_1\cdot p_2\right)^{\tilde  y_{12}}\left(\epsilon_3\cdot p_1\right)^{\tilde  y_{31}}\left(\epsilon_2\cdot p_1\right)^{\tilde  y_{21}}\, ,
\end{align}
where $D$ is an arbitrary constant and the exponents, non-negative integers, have to satisfy  
\begin{subequations}
\label{eq:onconsodd}
\begin{align}
\tilde x_{12}+\tilde x_{13}+\tilde y_{12}+z_3&=s_1\, , \\
\tilde x_{12}+\tilde x_{23}+\tilde y_{21}+z_4&=s_2\, , \\
\tilde x_{13}+\tilde x_{23}+\tilde y_{31}+z_5&=s_3\, , \\
z_1+z_2+z_3+z_4+z_5&=4\, ,\\
s_i\geq 0\, ,\\
1 \geq z_i\geq 0\, .
\end{align}
\end{subequations}
We can exploit the fact that we are working in $d=4$  to reduce the number of independent contributions. The contraction of the Levi-Civita tensor with five linearly dependent five-vectors $\{P_1,P_2,E_1,E_3,E_3\}$ yields
\begin{align}
\varepsilon\left(P_1,P_2,E_1,E_2,E_3\right)=0\, ,
\end{align}
where  we will take $P_1=\left(C_1^0,p_1^\mu\right)$, $P_2=\left(C_2^0,p_2^\mu\right)$, $E_1=\left(C_3^0,\epsilon_1^\mu\right)$, $E_2=\left(C_4^0,\epsilon_2^\mu\right)$, $E_3=\left(\sum\limits_{\nu=0}^4 \alpha_\nu C^0_\nu,\epsilon_3^\mu\right)$; $C_\mu^0$  are arbitrary scalars and the $\alpha_\nu$ are defined such that that $\epsilon_3^\mu=\alpha_0 p_1^\mu +\alpha_2 p_2^\mu+\alpha_3 \epsilon_1^\mu+\alpha_4 \epsilon_2^\mu$,  following the ideas used in \cite{Costa:2011mg, Bonifacio:2018vzv}. From the choices $\{C_1^0, C_2^0, C_3^0, C_4^0 \} = \{ B_{1j}, B_{2j}, A_{j1}, A_{j2} \}$ for $j=1,2,3$ and $\{C_1^0, C_2^0, C_3^0, C_4^0 \} = \{ A_{1j}, A_{2j}, p_j\cdot p_1 ,p_j\cdot p_2  \}$ for $j=1,2$ one can impose \cite{Bonifacio:2018vzv}
\begin{subequations}
\label{eq:oddcons}
\begin{align}
& B_{13} \varepsilon(p_1, p_2, \epsilon_1, \epsilon_2)-B_{12} \varepsilon(p_1, p_2,  \epsilon_1, \epsilon_3)-A_{12} \varepsilon(p_1, \epsilon_1, \epsilon_2, \epsilon_3)=0, \\
& B_{12} \varepsilon(p_1, p_2, \epsilon_2, \epsilon_3)+B_{23} \varepsilon(p_1, p_2, \epsilon_1, \epsilon_2)-A_{23} \varepsilon(p_2, \epsilon_1, \epsilon_2, \epsilon_3)=0, \\
& B_{13} \varepsilon(p_1, p_2, \epsilon_2, \epsilon_3)-B_{23} \varepsilon(p_1, p_2, \epsilon_1, \epsilon_3)+A_{31}\left(  \varepsilon(p_1, \epsilon_1, \epsilon_2, \epsilon_3)+\varepsilon(p_2, \epsilon_1, \epsilon_2, \epsilon_3)\right)=0, \\
& A_{23} \varepsilon(p_1, p_2, \epsilon_1, \epsilon_3)+ A_{31} \varepsilon(p_1, p_2, \epsilon_1, \epsilon_2)+p_{1}\cdot p_1 \varepsilon(p_2, \epsilon_1,  \epsilon_2, \epsilon_3)-p_{1}\cdot p_2 \varepsilon(p_1, \epsilon_1, \epsilon_2, \epsilon_3) =0, \\
& A_{12} \varepsilon(p_1, p_2, \epsilon_2,  \epsilon_3)- A_{31} \varepsilon(p_1, p_2,  \epsilon_1, \epsilon_2)+p_{1}\cdot p_2 \varepsilon(p_2, \epsilon_1,  \epsilon_2, \epsilon_3)-p_{2}\cdot p_2 \varepsilon(p_1, \epsilon_1,  \epsilon_2, \epsilon_3) =0.
\end{align}
\end{subequations}

In the case of massless particles of spin $\geq 1$, the functions \eqref{eq:amplieven} and \eqref{eq:ampliodd} must also be gauge invariant, which in this context means invariant under
\begin{align}
\label{eq:gaugin}
\epsilon_{i\mu}\rightarrow \epsilon_{i\mu}+ \alpha p_{i\mu}\, ,
\end{align}
 being $\alpha$ an arbitrary constant.

Finding all possible solutions to equations \eqref{eq:onconseven}  and \eqref{eq:onconsodd}, and considering only the amplitudes invariant under \eqref{eq:gaugin} if there are massless  particles with spin $\geq 1$, is equivalent to finding all possible on-shell three-point parity-even and parity-odd interactions, respectively. This procedure can be used for any three particles of any mass and spin. As a last step, if some particles are identical, we need to impose invariance under the interchange of these two particles, which reduces the list of allowed interactions.

Along section \ref{sec:threepoint} and tables \ref{MMM2}-\ref{M2M2M0} we  listed these solutions, considering three identical massive spin-2 particles and  two identical massive spin-2 particles coupled to a massless spin-2, massive spin-1 and massive or massless scalar particle. We only wrote the independent contributions in $d=4$  in the main text, showing below the full list of possibilities.

\subsection{On-shell three-point interactions, complete list}
\label{ap:pd3pfunc}

\subsubsection*{Massive-2, massive-2, massive-2}

We give in table \ref{MMM2ap} the list of allowed three-point interactions for this case.
\begin{table}[H]
\centering
\begin{tabular}{c }
$\mathcal{M}_{2,2,2}\left(m,m,m\right)$   \\ \hline
 $\mu B_{12}B_{13}B_{23}\equiv \mathcal{M}_1$	\\
  $\frac{1}{\Lambda_2}\left( B_{13}^2A_{21}^2+B_{23}^2A_{12}^2+B_{12}^2A_{31}^2\right)\equiv \mathcal{M}_2$	\\
 $\frac{1}{\Lambda_3}\left( B_{12}B_{23}A_{12}A_{32}+B_{12}B_{13}A_{21}A_{31}+B_{13}B_{23}A_{12}A_{21}\right)\equiv \mathcal{M}_3$	 \\
$\frac{1}{\Lambda_4^3} A_{31}A_{12}A_{21}\left(B_{23}A_{12}+B_{12}A_{31}-B_{13}A_{21}\right)\equiv \mathcal{M}_4$	\\
$\frac{1}{\Lambda_5^5}A_{12}^2A_{23}^2A_{31}^2\equiv \mathcal{M}_5$	 \\
$ \frac{1}{\Lambda_6}\left( B_{13}\, B_{23}\, \varepsilon \left( p_1,p_2,\epsilon_1,\epsilon_2\right)-B_{12}\, B_{23}\, \varepsilon \left( p_1,p_2,\epsilon_1,\epsilon_3\right)+B_{12}\, B_{13}\, \varepsilon \left( p_1,p_2,\epsilon_2,\epsilon_3\right)\right)\equiv \mathcal{M}_6$\\
$\frac{1}{\Lambda_7}\left(\left(A_{31} B_{12}-A_{12} B_{23} \right)\varepsilon\left(p_1, \epsilon_1, \epsilon_2, \epsilon_3\right)+\left(A_{31} B_{12} +A_{21}  B_{13} \right) \varepsilon\left(p_2, \epsilon_1, \epsilon_2, \epsilon_3\right)\right)\equiv  \mathcal{M}_7$\\
$\frac{1}{\Lambda_8^3} \left( A_{31}^2 B_{12}\varepsilon\left(p_1, p_2,  \epsilon_1, \epsilon_2\right)-A_{21}^2 B_{13} \varepsilon\left(p_1, p_2,  \epsilon_1, \epsilon_3\right)+A_{12}^2 B_{23} \varepsilon\left(p_1, p_2,  \epsilon_2, \epsilon_3\right)\right)\equiv  \mathcal{M}_8$\\
$\frac{1}{\Lambda_9^3}( A_{31}\left(-A_{21} B_{13} +A_{12}B_{23}\right)\varepsilon\left(p_1, p_2,  \epsilon_1, \epsilon_2\right)+A_{21}\left( A_{31} B_{12}
+A_{12} B_{23} \right)\varepsilon\left(p_1, p_2,  \epsilon_1, \epsilon_3\right)$\\ $+A_{12}\left(  A_{31} B_{12} - A_{21}
  B_{13}\right) \varepsilon\left(p_1, p_2,  \epsilon_2, \epsilon_3\right))\equiv\mathcal{M}_{9}$\\
$\frac{1}{\Lambda_{10}^5}\left( A_{12}A_{23}A_{31}\left(A_{31}\,\varepsilon \left( p_1,p_2,\epsilon_1,\epsilon_2\right)-A_{23}\,\varepsilon \left( p_1,p_2,\epsilon_1,\epsilon_3\right)+A_{12}\,\varepsilon \left( p_1,p_2,\epsilon_2,\epsilon_3\right)\right)\right)\equiv  \mathcal{M}_{10}$\\

	\end{tabular}
\caption{Three-point amplitudes for three identical massive  spin 2-particles with mass $m$.  } \label{MMM2ap}
\end{table}

Since we are working in $d=4$  any set of five or more vectors will always be linearly dependent -see the discussion in the previous section of this appendix-. In the parity-even sector this redundancy translates into the following relation, obtained by imposing the vanishing of the Gram matrix of the vectors $\{p_i,\epsilon_{i\mu}\}$ :
\begin{align}
\label{grammas}
-3 \mathcal{M}_1 \frac{m^4}{\mu}+2 \mathcal{M}_2\Lambda_2 m^2-2 \mathcal{M}_3 \Lambda_3 m^2+4 \Lambda_4^3\mathcal{M}_4=0\, .
\end{align}
This equation allows us to write, for instance, $\mathcal{M}_4$ as a function of the other amplitudes. The same can be done in the parity-odd sector, where from \eqref{eq:oddcons}  it follows \cite{Bonifacio:2018vzv}:
\begin{subequations}
\begin{align}
\mathcal{M}_9&=0\, ,\\
2\Lambda_6 \mathcal{M}_6+\Lambda_7\mathcal{M}_7&=0\, ,\\
3m^2 \Lambda_6\mathcal{M}_6+2\Lambda_8^3\mathcal{M}_8&=0\, ,
\end{align}
\end{subequations}
and so $\{ \mathcal{M}_7, \mathcal{M}_8, \mathcal{M}_9\}$ can be written as functions of $\{\mathcal{M}_6,\mathcal{M}_{10}\}$. A list with only the independent amplitudes  in four dimensions (and a  slightly different notation)was shown in table \ref{MMM2}.

\subsubsection*{Massive-2, massive-2, massless-2}

We start with six distinct parity-even and  nine parity-odd three-point operators, see table \ref{MM02a}
\begin{table}[H]
\centering
\begin{tabular}{c}
$\mathcal{G}_{2,2,2}\left(m,m,0\right)$   \\  \hline
$\frac{1}{M_p} A_{31}^2 B_{12}^2\equiv \mathcal{G}_1$ \\
$\frac{1}{M_p} A_{31} B_{12} \left(A_{12} B_{23}+A_{23} B_{13}\right)\equiv \mathcal{G}_2$  \\
$\frac{1}{M_p}\left(A_{23} B_{13}+A_{12} B_{23}\right){}^2\equiv \mathcal{G}_3$	\\
$\frac{1}{M_p \hat\Lambda_4^2}A_{12} A_{23} A_{31} \left(A_{12} B_{23}+A_{23} B_{13}\right)\equiv \mathcal{G}_4$ \\
$\frac{1}{M_p \hat\Lambda_5^2}A_{12} A_{21} A_{31}^2 B_{12}\equiv \mathcal{G}_5$\\
$\frac{1}{M_p \hat{\Lambda}_6^4}A_{12}^2 A_{23}^2 A_{31}^2\equiv \mathcal{G}_6$	 \\
$\frac{1}{M_p}\left(B_{23}A_{12}+B_{13}A_{23}\right)\,\varepsilon\left(p_3 ,\epsilon_1 , \epsilon_2 , \epsilon_3\right)\equiv \mathcal{G}_7$\\
$\frac{1}{M_p}B_{12}A_{31}\,\varepsilon\left(p_3 ,\epsilon_1 , \epsilon_2 , \epsilon_3\right)\equiv \mathcal{G}_8$\\
$\frac{1}{M_p\hat{\Lambda}_9^2}A_{12}A_{23}A_{31}\,\varepsilon\left(p_3 ,\epsilon_1, \epsilon_2 , \epsilon_3\right)\equiv \mathcal{G}_9$\\
$\frac{1}{M_p\hat{\Lambda}_{10}^2}A_{31} B_{12} \left(A_{21} \varepsilon\left(p_1 , p_2 ,\epsilon_1\, \epsilon_3\right)+A_{12} \varepsilon\left(p_1 , p_2 ,\epsilon_2\, \epsilon_3\right)\right)\equiv \mathcal{G}_{10}$\\
$\frac{1}{M_p\hat{\Lambda}_{11}^2}A_{31}^2 B_{12}\varepsilon\left(p_1 , p_2 ,\epsilon_1\, \epsilon_2\right)\equiv\mathcal{G}_{11}$\\
$\frac{1}{M_p\hat{\Lambda}_{12}^2}\left(A_{12} B_{23}-A_{21} B_{13}\right) \left(A_{21}\varepsilon\left(p_1 , p_2 ,\epsilon_1\, \epsilon_3\right)+A_{12}\varepsilon\left(p_1 , p_2 ,\epsilon_2\, \epsilon_3\right) \right)\equiv\mathcal{G}_{12}$\\
$\frac{1}{M_p\hat{\Lambda}_{13}^2}A_{31} \left(A_{12} B_{23}-A_{21} B_{13}\right)\varepsilon\left(p_1 , p_2 ,\epsilon_1\, \epsilon_2\right) \equiv\mathcal{G}_{13}$\\
$\frac{1}{M_p\hat{\Lambda}_{14}^4}A_{21} A_{31} A_{12}\left(A_{21}\varepsilon\left(p_1 , p_2 ,\epsilon_2\, \epsilon_3\right) + A_{12}\varepsilon\left(p_1 , p_2 ,\epsilon_1\, \epsilon_3\right)\right)\equiv\mathcal{G}_{14} $\\
$\frac{1}{M_p\hat{\Lambda}_{15}^4}A_{12}A_{23}A_{31}^2\,\varepsilon\left(p_1 , p_2 ,\epsilon_1\, \epsilon_2\right)\equiv \mathcal{G}_{15}$
\end{tabular}
\caption{All possible three-point amplitudes for   two (identical) massive spin-2 particles, one massless  spin 2-particle. } \label{MM02a}
\end{table}

The fact that we are working in $d=4$ implies, for the parity-even sector:
\begin{align}
m^2\mathcal{G}_3+2\hat{\Lambda}_4^2\mathcal{G}_4+2 \hat{\Lambda}_5^2\mathcal{G}_5=0\, ,
\end{align}
which we will use to rewrite $\mathcal{G}_4$ as a function of the other $\mathcal{G}_i$. On the other hand, the parity-odd sector can be reduced by using \eqref{eq:oddcons}, which translates into
\begin{subequations}
\begin{align}
\hat{\Lambda}_{14}^4\mathcal{G}_{14}+2\hat{\Lambda}_{15}^2m^2\mathcal{G}_{9}+2\hat{\Lambda}_{15}^4 \mathcal{G}_{15}&=0 \, ,\\
 \hat{\Lambda}_{12}^2\mathcal{G}_{12}+2 \hat{\Lambda}_{9}^2\mathcal{G}_{9}&=0\, ,\\
  \hat{\Lambda}_{12}^2\mathcal{G}_{12}-  \hat{\Lambda}_{13}^2\mathcal{G}_{13}+ \hat{\Lambda}_{9}^2 \mathcal{G}_{9}+m^2 \mathcal{G}_{7}&=0\, ,\\
   \hat{\Lambda}_{13}^2\mathcal{G}_{13}+  \hat{\Lambda}_{10}^2\mathcal{G}_{10}&=0\\
 \hat{\Lambda}_{10}^2\mathcal{G}_{10}-2 \hat{\Lambda}_{11}^2\mathcal{G}_{11}-2m^2\mathcal{G}_{8} &=\, ,0
\end{align}
\end{subequations}
so we can also ignore the set $\{ \mathcal{G}_{10},\mathcal{G}_{11},\mathcal{G}_{12},\mathcal{G}_{13},\mathcal{G}_{14}\}$ as long as we work in four dimensions. This is what we did in table \ref{MM02}.

\subsubsection*{Massive-2, massive-2, massive-1}

The list of possible three-point functions, two parity-even and five parity-odd, is given in table \ref{M2M2M1a}.
\begin{table}[H]
\centering
\begin{tabular}{c}
$\mathcal{M}_{2,2,1}\left(m,m,m_1\right)$  \\ \hline
$\left( A_{12} B_{12} B_{23}+A_{21} B_{12} B_{13}\right)\equiv \mathcal{K}_1$\\
 $\frac{1}{\tilde{\Lambda}_2^2}\left( A_{21} A_{12}^2 B_{23}+A_{21}^2 A_{12} B_{13}\right)\equiv \mathcal{K}_2$\\
$B_{12}\left(\varepsilon\left(p_1,\epsilon_1,\epsilon_2,\epsilon_3\right)-\varepsilon\left(p_2,\epsilon_1,\epsilon_2,\epsilon_3\right)\right)\equiv \mathcal{K}_3$\\
$\frac{1}{\tilde{\Lambda}_4^2}\, \varepsilon\left(p_1,p_2,\epsilon_1,\epsilon_2\right)\left(B_{23}A_{12}-B_{13}A_{23}\right)\equiv \mathcal{K}_4$\\
$\frac{1}{\tilde{\Lambda}_5^2}A_{12} A_{21} (\left(p_2,\epsilon_1,\epsilon_2,\epsilon_3\right)-\left(p_1,\epsilon_1,\epsilon_2,\epsilon_3\right))\equiv\mathcal{K}_5$\\
$\frac{1}{\tilde{\Lambda}_6^2}B_{12} \left(A_{21}\left(p_1,p_2,\epsilon_1,\epsilon_3\right) -A_{12}\left(p_1,p_2,\epsilon_2,\epsilon_3\right)\right)\equiv\mathcal{K}_6$\\
$\frac{1}{\tilde{\Lambda}_7^4} A_{12} A_{21} \left(A_{21} \left(p_1,p_2,\epsilon_1,\epsilon_3\right)-A_{12} \left(p_1,p_2,\epsilon_2,\epsilon_3\right)\right)\equiv\mathcal{K}_7$
	\\\end{tabular}
\caption{All possible three-point amplitudes for two identical  massive  spin-2 particles and one massive spin-1 particle.} \label{M2M2M1a}
\end{table}
The parity-even sector cannot be reduced further by using dimensional dependent relations. On the other hand, by imposing \eqref{eq:oddcons}, in four dimensions the parity-odd terms satisfy
\begin{subequations} 
\begin{align}
\mathcal{K}_5+\mathcal{K}_4-\mathcal{K}_6&=0\, ,\\
2\mathcal{K}_6-m_1^2\mathcal{K}_3&=0\, ,\\
2\mathcal{K}_7+m_1^2\mathcal{K}_5&=0\, ,
\end{align}
\end{subequations}
which allows us to eliminate $\{\mathcal{K}_5,\mathcal{K}_6,\mathcal{K}_7\}$ in favour of $\{\mathcal{K}_3, \mathcal{K}_4\}$. This is how we presented the results in the main text, see table \ref{M2M2M1}.

\section{Lagrangian basis}
\label{ap:lagbasis}

As explained in the main text, one of the advantages of working directly with on-shell amplitudes is that they are blind to field redefinitions and integration by parts in the Lagrangian, making the correspondence between both pictures  not one-to-one. In this appendix we will give a Lagrangian basis for the  parity-even amplitudes presented in  section \ref{sec:ampli} and appendix \ref{ap:pd3pfunc}, focusing only on the three-point amplitudes. Notice that Lagrangians are off-shell quantities, and only when going on-shell one reproduces the previous results. This means that when we write one of the bases as a linear combination of the other one, both sides must  be read on-shell.

\textbf{Conventions and definitions}

The Riemann tensor is:
\begin{align}
R_{\alpha \beta \mu \nu}=\frac{1}{2}\left[\partial_\mu \partial_\beta h_{\nu \alpha}+\partial_\nu \partial_\alpha h_{\beta \mu}-\partial_\nu \partial_\beta h_{\mu \alpha}-\partial_\mu \partial_\alpha h_{\beta \nu}\right]\, , 
\end{align}
and we define
\begin{equation}
F_{\alpha\beta\mu}= \partial_\alpha h_{\beta\mu}-\partial_\beta h_{\alpha\mu}\, .
\end{equation}
where $h_{\mu\nu}$ represents the massive spin-2 field. Instead of $h_{\mu\nu}$ and their derivatives we are using a very particular linear combination of them defined as $F_{\alpha\beta\nu}$ and $R_{\alpha\beta\mu\nu}$. These fields help us capture a particular polarization of the massive spin-2 particle. For example, in Regge limit $h_{\mu\nu}$ may have all possible polarizations but roughly speaking, at leading order $R_{\alpha\beta\mu\nu}$ captures only the transverse polarization and filters out the vector and the longitudinal polarizations. Similarly $F_{\alpha\beta\nu}$ captures vector polarization at leading order. To go from the derivative basis $\partial_\mu$ to the momentum basis $p_\mu$ we do $\partial_\mu\rightarrow -i p_\mu$. We denote by $\mathcal{L}_{a,b,c}\left(m_a,m_b,m_c\right)$ a three-point interaction of particles   with spin $a$, $b$, $c$ and masses $m_a$, $m_b$, $m_c$, where the particle $c$ will be the one exchanged. Indices are contracted with the flat metric $\eta_{\mu\nu}$.

\subsection*{Massive-2, massive-2, massive-2}
\label{ap:submasmasimasi}

A Lagrangian basis for the on-shell parity-even three-point amplitudes listed in table \ref{MMM2ap} is
\begin{table}[H]
\centering
\begin{tabular}{c}
	Lagrangian basis $\mathcal{L}_{2,2,2}\left(m,m,m\right)$ \\ \hline
	$\bk{-}\frac{1}{\Lambda_{ L_{\hat 5}}^5}\left( R^{1\mu\nu}_{\alpha\beta}R^{2\gamma\delta}_{\mu\nu}R^{3\alpha\beta}_{\gamma\delta}\right)\equiv L_{\hat 5}$\\
	$\frac{1}{\Lambda_{ L_{\hat 4}}^3}\left( R^{1\mu\nu\alpha\beta}F^{2\delta}_{\mu\nu}F^3_{\alpha\beta\delta}+F^{1\delta}_{\mu\nu}R^{2\mu\nu\alpha\beta}F^3_{\alpha\beta\delta}+F^1_{\alpha\beta\delta}F^{2\delta}_{\mu\nu}R^{3\mu\nu\alpha\beta}\right)\equiv L_{\hat 4}$\\	
	$\bk{-}\frac{1}{\Lambda_{ L_{\hat 3}}}\left(F^{1\mu\alpha}_{\beta}F^2_{\nu\mu\alpha}h^{3\beta\nu}+F^{1\mu\alpha}_{\beta}h^{2\beta\nu}F^3_{\nu\mu\alpha}+h^{1\beta\nu}F^{2\mu\alpha}_{\beta}F^3_{\nu\mu\alpha}\right)\equiv L_{\hat 3}$\\
	$\bk{-}\frac{1}{\Lambda_{ L_{\hat 2}}}\left( h^{1\mu\alpha}h^{2\nu\beta}R^3_{\mu\nu\alpha\beta}+h^{1\mu\alpha}R^2_{\mu\nu\alpha\beta}h^{3\nu\beta}+R^1_{\mu\nu\alpha\beta}h^{2\nu\beta}h^{3\mu\alpha}\right)\equiv L_{\hat 2}$\\
	$\mu\left( h^{1\nu}_{\mu} h^{2\alpha}_{\nu}h^{3\mu}_{\alpha}\right)\equiv L_{\hat 1}$\\
	\end{tabular}
\caption{Lagrangians term for the three-point vertices derived in table \ref{MMM2ap}} \label{MMM2l}
\end{table}
where $\Lambda_{L\hat i}$ and $\mu$ have units of energy. On-shell and Lagrangian basis are related by
\begin{subequations}
\label{amla2m}
\begin{align}
\mathcal{M}_1&= L_{\hat 1}\, , \\
-2\mathcal{M}_2&= 2L_{\hat 2}-2L_{\hat 3}+3m^2L_{\hat 1}\, ,\\
-2\mathcal{M}_3&=L_{\hat 2}-2L_{\hat 3}+3m^2L_{\hat 1}\, , \\
4\mathcal{M}_4&=L_{\hat 4}+2m^2(L_{\hat 2}-2L_{\hat 3}+3m^2L_{\hat 1})+3m^4 L_{\hat 1}\, , \\
-8\mathcal{M}_5&=L_{\hat 5}-m^2L_{\hat 4}-m^4\left(L_{\hat 2}-2L_{\hat 3}\right)-5m^6 L_{\hat 1}\, .\nonumber
\end{align}
\end{subequations}
where we are doing an abuse of notation, to maintain the expressions simple, calling $\frac{\mathcal{M}_1}{\mu}\equiv \mathcal{M}_1,\,  \Lambda_{L\hat{i}}\mathcal{M}_i\equiv \mathcal{M}_i$, $\frac{L_{\hat 1}}{\mu}\equiv L_{\hat 1},\, \Lambda_{L\hat{i}}L_{\hat i}\equiv L_{\hat i}$. In other words,  $\mathcal{M}_i$ and $L_{\hat i}$  in \eqref{amla2m} do not necessarily have mass dimension=4.

\subsection*{Massive-2, massive-2, massless-2}
\label{ap:submasmasimass}

An off-shell basis for the six  parity-even contributions presented in table \ref{MM02a} is
\begin{table}[H]
\centering
\begin{tabular}{ c}
 Lagrangian basis  $\mathcal{L}_{2,2,2}\left(m,m,0\right)$ \\  \hline
$\bk{-}\frac{1}{\hat{\Lambda}_{L6}^4}\left( R^{1\mu\nu}_{\alpha\beta}R^{2\gamma\delta}_{\mu\nu}R^{3\alpha\beta}_{\gamma\delta}\right)\equiv L_6$  \\
 $\frac{1}{\hat{\Lambda}_{L5}^2}\left( F^{1\gamma}_{\mu\nu}F^2_{\alpha\beta\gamma}R^{3\mu\nu\alpha\beta}\right)\equiv L_5$	\\
 $\frac{1}{\hat{\Lambda}_{L4}^2}\left( R^{1\mu\nu\alpha\beta}F^{2\delta}_{\mu\nu}F^3_{\alpha\beta\delta}+F^{1\delta}_{\mu\nu}R^{2\mu\nu\alpha\beta}F^3_{\alpha\beta\delta}\right)\equiv L_4$\\
 $\bk{-}\left( h^1_{\mu\alpha}h^2_{\beta\nu}R^{3\mu\nu\alpha\beta}\right)\equiv L_3$\\
 $\bk{-}\left( F^{1\alpha}_{\beta\mu}h^{2\beta\nu}F^3_{\nu\mu\alpha}+h^{1\beta\nu}F^{2\alpha}_{\beta\mu}F^3_{\nu\mu\alpha}\right)\equiv L_2$  \\
  $\bk{-}\left( F^1_{\beta\mu\alpha}F^{2\mu\alpha}_{\nu}h^{3\beta\nu}-\bk{\partial_\gamma h^{1\mu}_{\nu}\partial^\gamma h^{2\alpha}_{\nu}h^{3\mu}_{\alpha}}\right)\equiv L_1$ \\
\end{tabular}
\caption{Lagrangian basis for the parity-even three-point amplitudes of   two (identical) massive spin-2 particles, one massless  spin 2-particle. } \label{MM02L}
\end{table}
where, as usual, the $\hat{\Lambda}_{Li}$ have mass dimension=1. Both bases are related  (again making an abuse of notation and calling $\hat \Lambda_i\mathcal{G}_i\equiv \mathcal{G}_i$, $\hat\Lambda_{Li}L_i\equiv L_i$) by:
\begin{subequations}
\begin{align}
-8\mathcal{G}_6&=L_6-2m^2 L_5 \, , \\
4\mathcal{G}_5&=L_5 \, , \\
4\mathcal{G}_4&=L_4-4m^2(L_2 -L_3) \, , \\
-\mathcal{G}_3&=L_3 \, , \\
-\mathcal{G}_2&=L_2 -L_3  \, , \\
-\mathcal{G}_1&=L_1-L_2 +L_3 \, .
\end{align}
\end{subequations}

\subsection*{Massive-2, massive-2, massive-1}
\label{ap:submasmasimasi1}

Regarding table \ref{M2M2M1}, the parity-even entries can be described by the following Lagrangians
\begin{table}[H]
\centering
\begin{tabular}{c}
Lagrangian basis $\mathcal{L}_{2,2,1}\left(m,m,m_1\right)$ \\ \hline
$\frac{1}{\tilde{\Lambda}_L^2}\left( R^{1\alpha\beta\mu\nu}F^2_{\alpha\beta\mu}A^3_\nu+F^1_{\alpha\beta\mu}R^{2\alpha\beta\mu\nu}A^3_\nu\right)\equiv L_{\tilde 2}$\\
 $F^1_{\mu\alpha\beta} h^{2\alpha\beta}A^{3\mu}+ h^{1\alpha\beta}F^2_{\mu\alpha\beta}A^{3\mu}\equiv L_{\tilde 1}$\\
	\\\end{tabular}
\caption{Lagrangians describing parity-even  three-point amplitudes for two identical  massive  spin-2 particles and one massive spin-1 particle} \label{M2M2M1l}
\end{table}
where $\tilde{\Lambda}_L^2$ is some energy scale.  Both languages are related (abusing of notation and calling $\tilde \Lambda_2^2\mathcal{K}_2\equiv \mathcal{K}_2$, $\tilde{\Lambda}_L^2L_{\tilde 2}\equiv L_{\tilde 2}$)) by 
\begin{subequations}
\begin{align}
\bk{i}\mathcal{K}_1&=L_{\tilde 1}\, ,\\
2\bk{i}\mathcal{K}_2&=L_{\tilde 2}\bk{+}2\Delta_{m_1} L_{\tilde 1}\, ,
\end{align}
\end{subequations}
where $\Delta_{m_1}=\frac{2m^2-m_1^2}{2}$.

\subsection*{Massive-2, massive-2,scalar}
\label{ap:submasmasscalar}

Finally, the parity-even  part of table \ref{M2M2M0} can be obtained from
\begin{table}[H]
\centering
\begin{tabular}{ c}
Lagrangian basis $\mathcal{L}_{2,2,0}\left(m,m,M\right)$  \\ \hline
$\frac{1}{\Lambda_{L_{\dot 3}}^3}\left( R^{1\mu\nu\alpha\beta}R^2_{\mu\nu\alpha\beta}\phi\right)\equiv L_{\dot 3}$\\
$\frac{1}{\Lambda_{L_{\dot 2}}}\left( F^{1\mu\nu\alpha}F^2_{\mu\nu\alpha}\phi\right)\equiv L_{\dot 2}$\\
 $m_s\left( h_{\mu\nu}h^{\mu\nu}\phi\right)\equiv L_{\dot 1}$\\
	\\\end{tabular}
\caption{Lagrangians for two  massive  spin 2-particles and one scalar particle} \label{M2M2M0l}
\end{table}
with $\Lambda_{L_{\dot i}}$ an energy scale supressing the non-renormalizable operators. On-shell and Lagrangian pictures (abusing of notation and calling $ \frac{\mathcal{S}_1}{m_s}\equiv \mathcal{S}_1,\,\Lambda_{L_{\dot i}}{\mathcal{S}_i}\equiv \mathcal{S}_i, \frac{L_{\dot 1}}{m_s}\equiv L_{\dot 1},\,  \Lambda_{L_{\dot i}}L_{\dot i}\equiv L_{\dot i}$) are related by
\begin{subequations}
\begin{align}
4\mathcal{S}_3&=L_{\dot 3}+4\Delta_M^2L_{\dot 1}-4i\Delta_M L_{\dot 2}\, , \\
-2\mathcal{S}_2&=iL_{\dot 2}-2\Delta_M L_{\dot 1}\,\\ 
\mathcal{S}_1&=L_{\dot 1}\, , 
\end{align}
\end{subequations}
where $\Delta_{M}=\frac{2m^2-M^2}{2}$.

\section{Fierz-Pauli vertices}
\label{ap:fpver}

In this appendix we will derive the contribution $\nabla_\mu h_{\rho\sigma} \nabla^\mu h^{\rho\sigma}$. The term $\nabla_\mu h^{}_{\rho\sigma} \nabla^\rho h^{\mu\sigma}$ can be obtained in a similar way. For simplicity we will take  $M_p\equiv 1$  in the computation.

\begin{align}
&\tilde g^{\mu\nu}\tilde g^{\rho\alpha}\tilde g^{\sigma\beta}\nabla_\mu h_{\rho\sigma}\nabla_\nu h_{\alpha\beta}|_{hhg}=\nonumber\\&\left(\eta^{\mu\nu}-g^{\mu\nu}\right)\left(\eta^{\rho\alpha}-g^{\rho\alpha}\right)\left(\eta^{\sigma\beta}-g^{\sigma\beta}\right)\left(\partial_\mu h_{\rho\sigma}-\Gamma^\gamma_{\mu \rho}h_{\gamma\sigma}-\Gamma^\gamma_{\mu \sigma}h_{\rho\gamma}\right)\left(\partial_\nu h_{\alpha\beta}-\Gamma^\gamma_{\nu_\alpha}h_{\gamma\beta}-\Gamma^\gamma_{\nu_\beta}h_{\alpha\gamma}\right)|_{hhg}=\nonumber\\&-\left(g^{\mu\nu}\eta^{\rho\alpha}\eta^{\sigma\beta}+\eta^{\mu\nu}g^{\rho\alpha}\eta^{\sigma\beta}+\eta^{\mu\nu}\eta^{\rho\alpha}g^{\sigma\beta}\right)\partial_\mu h_{\rho\sigma}\partial_\nu h_{\alpha\beta}\nonumber\\&+\eta^{\mu\nu}\eta^{\rho\alpha}\eta^{\sigma\beta}\left(\partial_\mu h_{\rho\sigma}-\Gamma^\gamma_{\mu \rho}h_{\gamma\sigma}-\Gamma^\gamma_{\mu \sigma}h_{\rho\gamma}\right)\left(\partial_\nu h_{\alpha\beta}-\Gamma^\gamma_{\nu_\alpha}h_{\gamma\beta}-\Gamma^\gamma_{\nu_\beta}h_{\alpha\gamma}\right)|_{hhg}\, .
\end{align}
Let us write explicitly the contribution $\left(\Gamma^\gamma_{\mu \rho}h_{\gamma\sigma}\right)|_{hhg}$
\begin{align}
\Gamma^\gamma_{\mu\rho}h_{\gamma\sigma}|_{hhg}=\frac{1}{2}\eta^{\gamma\lambda}\left(\partial_{\mu}g_{\rho\lambda}+\partial_{\rho}g_{\mu\lambda}-\partial_{\lambda}g_{\mu\rho}\right)h_{\gamma\sigma}\, .
\end{align}
Expanding and going on-shell
\begin{align}
&+\eta^{\mu\nu}\eta^{\rho\alpha}\eta^{\sigma\beta}\left(\partial_\mu h_{\rho\sigma}-\Gamma^\gamma_{\mu \rho}h_{\gamma\sigma}-\Gamma^\gamma_{\mu \sigma}h_{\rho\gamma}\right)\left(\partial_\nu h_{\alpha\beta}-\Gamma^\gamma_{\nu_\alpha}h_{\gamma\beta}-\Gamma^\gamma_{\nu_\beta}h_{\alpha\gamma}\right)|_{hhg}=\nonumber\\ & -\frac{1}{2}\eta^{\gamma\lambda}\eta^{\mu\nu}\eta^{\rho\alpha}\eta^{\sigma\beta}\left(\partial_{\mu}g_{\rho\lambda}+\partial_{\rho}g_{\mu\lambda}-\partial_{\lambda}g_{\mu\rho}\right)h_{\gamma\sigma}\partial_\nu h_{\alpha\beta}\nonumber \\ &-\frac{1}{2}\eta^{\mu\nu}\eta^{\rho\alpha}\eta^{\sigma\beta}\eta^{\gamma\lambda}\left(\partial_{\mu}g_{\sigma\lambda}+\partial_{\sigma}g_{\mu\lambda}-\partial_{\lambda}g_{\mu\sigma}\right)h_{\gamma\rho}\partial_\nu h_{\alpha\beta}\nonumber\\ & -\frac{1}{2}\eta^{\gamma\lambda}\eta^{\mu\nu}\eta^{\rho\alpha}\eta^{\sigma\beta}\left(\partial_{\mu}g_{\rho\lambda}+\partial_{\rho}g_{\mu\lambda}-\partial_{\lambda}g_{\mu\rho}\right)h_{\gamma\sigma}\partial_\nu h_{\alpha\beta}\nonumber \\ &-\frac{1}{2}\eta^{\mu\nu}\eta^{\rho\alpha}\eta^{\sigma\beta}\eta^{\gamma\lambda}\left(\partial_{\mu}g_{\sigma\lambda}+\partial_{\sigma}g_{\mu\lambda}-\partial_{\lambda}g_{\mu\sigma}\right)h_{\gamma\rho}\partial_\nu h_{\alpha\beta}=\nonumber \\ &\left(A_{23}A_{32}B_{13}B_{12}-A_{32}A_{13}B_{32}B_{12}\right)+\left(A_{13}A_{31}B_{23}B_{12}-A_{23}A_{31}B_{13}B_{23}\right)\nonumber\\&=-2A_{31}B_{12}\left(A_{23}B_{13}+A_{12}B_{23}\right)=-2\mathcal{M}_2\, ,
\end{align}
whereas the other contribution is just
\begin{align}
-\left(g^{\mu\nu}\eta^{\rho\alpha}\eta^{\sigma\beta}+\eta^{\mu\nu}g^{\rho\alpha}\eta^{\sigma\beta}+\eta^{\mu\nu}\eta^{\rho\alpha}g^{\sigma\beta}\right)\partial_\mu h_{\rho\sigma}\partial_\nu h_{\alpha\beta}=-A_{31}^2B_{12}^2+2m^2B_{12}B_{23}B_{13}\, .
\end{align}

\clearpage

\section{Contact terms}
\label{ap:contactterms}

To simplify the notation we will define  $A_1\equiv A_{12} =\epsilon_{1}\cdot p_2$, $A_2\equiv A_{21}= \epsilon_{2}\cdot p_1$, $A_3\equiv  A_{31}= \epsilon_{3}\cdot  p_1$, $A_4\equiv  A_{41}= \epsilon_{4}\cdot p_1$, $A_5\equiv  A_{14}= \epsilon_{1}\cdot p_4$, $A_6\equiv  A_{24}= \epsilon_{2}\cdot p_4$, $A_7\equiv  A_{34}= \epsilon_{3}\cdot p_4$, $A_8\equiv  A_{42}= \epsilon_{4}\cdot p_2$. We will show here the  parity-even tensor structures invariant under the kinetic permutations and  with up to two derivatives. The rest of the terms can be constructed using the attached  \texttt{Mathematica} notebook.
\subsubsection*{ With no derivatives}

For this case a Lagrangian basis is
\begin{table}[H]
\begin{center}
\begin{tabular}{ c| c } 
$\{A, B\}$ formalism	& Lagrangian basis\\
 \hline
 $B_{13} B_{14} B_{23} B_{24}$& $h^{1\mu\nu}h^{3\alpha}_\nu h^{2\beta}_\alpha h^{4}_{\beta\mu}$ \\
$B_{12} B_{14} B_{23} B_{34}$  &$h^{1\mu\nu}h^{2\alpha}_\nu h^{3\beta}_\alpha h^{4}_{\beta\mu}$  \\
$B_{12} B_{13} B_{24} B_{34}$ & $h^{1\mu\nu}h^{2\alpha}_\nu h^{4\beta}_\alpha h^{3}_{\beta\mu}$ \\
 $B_{14}^2 B_{23}^2$   & $h^{1\mu\nu} h^4_{\mu\nu}\, h^{2\alpha\beta} h^3_{\alpha\beta}$ \\
  $B_{13}^2 B_{24}^2$   & $ h^{1\mu\nu} h^3_{\mu\nu}\, h^{2\alpha\beta} h^4_{\alpha\beta}$ \\
   $B_{12}^2 B_{34}^2$   &$h^{1\mu\nu} h^2_{\mu\nu}\, h^{3\alpha\beta} h^4_{\alpha\beta}$  \\
\end{tabular}
\caption{\label{tab:cont0basis} Explicit basis for the parity-even tensor structures invariant under kinetic permutations and with 0 derivatives }
\end{center}
\end{table}
\subsubsection*{ With two derivatives}

We will only write the first element of each term, hiding in $\dots$ the elements needed to make the term invariant under the kinetic permutations -see the definition in expression \eqref{eq:permutationsym}-, e.g. $A_1 A_2 B_{34}^2 B_{12}+\dots\equiv A_1 A_2 B_{34}^2 B_{12}-A_4 A_7 B_{34} B_{12}^2-A_7 A_8 B_{34} B_{12}^2$. The basis, which has 30 terms, is 
\begin{table}[h!]
\centering
\small	
\begin{tabular}{ c|c || c|c }
$\{A, B\}$ formalism	& Lagrangian basis&$\{A, B\}$ formalism	& Lagrangian basis\\
\hline
$ A_4 A_8 B_{12} B_{13} B_{23}+\rm \dots$	&	 $\partial^\gamma h^{1\mu\nu}\partial^\beta h^{2\alpha}_\mu h^{3}_{\nu\alpha} h^4_{\gamma\beta}+\rm \dots$	& $A_5 A_1 B_{23} B_{24} B_{34}+\rm \dots$	&	 $\partial^\gamma h^{2\mu\nu} h^{3\alpha}_\mu \partial^\beta h^{4}_{\nu\alpha} h^1_{\gamma\beta}+\dots$\\ \hline
$ A_6 A_2 B_{13} B_{14} B_{34}+\dots$	&	$\partial^\gamma h^{1\mu\nu} h^{3\alpha}_\mu \partial^\beta h^{4}_{\nu\alpha} h^2_{\gamma\beta}+\dots$	&	$A_5 A_6
   B_{34}^2 B_{12}$	&	$h^{1\mu\nu} h^2_{\mu\gamma} h^{3\alpha\beta} \partial^\gamma \partial_\nu h^{4}_{\alpha\beta} +\dots$ \\\hline
$A_1 A_2 B_{34}^2 B_{12}+\dots$	&	$\partial^\gamma h^{1\mu\nu} \partial_\nu h^2_{\mu\gamma} h^{3\alpha\beta}   h^{4}_{\alpha\beta} +\dots$	&	$A_1 A_2 B_{13}
   B_{34} B_{24}+\dots$	&	$\partial^\alpha h^{1\mu\nu}h^{3}_{\mu\gamma}h^{4\gamma\beta}\partial_\nu  h^{2}_{\beta\alpha}+\rm \dots$ \\\hline
$A_1 A_2 B_{14}B_{34} B_{23}+\dots$	&	$\partial^\alpha h^{1\mu\nu}h^{4}_{\mu\gamma}h^{3\gamma\beta}\partial_\nu  h^{2}_{\beta\alpha}+\rm \dots$	&	$A_5 A_6 B_{13} B_{34} B_{24}+\dots$	&	$ h^{1\mu\nu}h^{3}_{\mu\gamma}\partial_\nu  \partial^\alpha h^{4\gamma\beta} h^{2}_{\beta\alpha}+\rm \dots$ \\\hline
$ A_5 A_6 B_{14} B_{34} B_{23}+\dots$	&	$h^{1\mu\nu}\partial_\nu  \partial^\alpha h^{4}_{\mu\gamma} h^{3\gamma\beta} h^{2}_{\beta\alpha}+ \dots$	&	$A_1 A_6 B_{34}^2 B_{12}+\dots$	&	$\partial^\gamma  h^{1\mu\nu} h^2_{\mu\gamma} h^{3\alpha\beta} \partial_\nu h^{4}_{\alpha\beta} +\dots$  \\\hline
$A_1
   A_6 B_{13} B_{34} B_{24}+\dots$	&	$ \partial^\alpha h^{1\mu\nu}h^{3}_{\mu\gamma}\partial_\nu   h^{4\gamma\beta} h^{2}_{\beta\alpha}+ \dots$	&	$A_1
   A_6 B_{14} B_{34} B_{23}+\dots$	&	$ \partial^\alpha h^{1\mu\nu}\partial_\nu  h^{4}_{\mu\gamma}  h^{3\gamma\beta} h^{2}_{\beta\alpha}+ \dots$ \\\hline
$A_2 A_8 B_{24} B_{13}^2+\dots$	&	$\partial^\gamma  h^{2\mu\nu} h^4_{\mu\gamma}  \partial_\nu h^{1\alpha\beta} h^{3}_{\alpha\beta} +\dots$	&	$A_2 A_8 B_{12} B_{13} B_{34}+\dots$	&	$\partial^\beta h^{1\mu\nu}\ h^{3\alpha}_\mu \partial^\gamma  h^{2}_{\nu \beta} h^4_{\gamma\alpha}+\rm \dots$	\\\hline
$A_2 A_8 B_{13} B_{14} B_{23}+\dots$	&	$\partial^\gamma  h^{1\mu\nu}\ h^{3\alpha}_\mu  h^{4}_{\nu \beta} \partial^\beta h^2_{\gamma\alpha}+\rm \dots$	&	$A_1 A_7 B_{24}^2 B_{13}+\dots$	&	$h^{1\mu\nu} h^3_{\mu\gamma}\partial_\nu h^{2\alpha\beta} \partial^\gamma  h^{4}_{\alpha\beta} +\dots$  \\\hline
$A_1 A_7 B_{12}
   B_{24} B_{34}+\dots$	&	$h^{1\mu\nu}\partial_\nu   h^{2}_{\mu\gamma} \partial^\alpha h^{4\gamma\beta} h^{3}_{\beta\alpha}+ \dots$	&	$A_1 A_7 B_{14}
   B_{24} B_{23}+\dots$	&	$h^{1\mu\nu} \partial^\alpha  h^{4}_{\mu\gamma}  \partial_\nu h^{2\gamma\beta} h^{3}_{\beta\alpha}+ \dots$ \\\hline
$A_5 A_7
   B_{24}^2 B_{13}+\dots$	&	$h^{1\mu\nu} h^3_{\mu\gamma}h^{2\alpha\beta} \partial^\gamma  \partial_\nu  h^{4}_{\alpha\beta} +\dots$	&	$A_3 A_5 B_{12} B_{24} B_{34}+\dots$	&	$\partial^\alpha h^{1\mu\nu}   h^{2}_{\mu\gamma}  \partial_\nu h^{4\gamma\beta} h^{3}_{\beta\alpha}+ \dots$ \\\hline
$ A_3 A_5 B_{14} B_{24} B_{23}+\dots$	&	$\partial^\alpha h^{1\mu\nu}  \partial_\nu   h^{4}_{\mu\gamma}  h^{2\gamma\beta} h^{3}_{\beta\alpha}+ \dots$ 	&	$A_2 A_7 B_{23} B_{14}^2+\dots$	&	$h^{2\mu\nu} h^3_{\mu\gamma}\partial_\nu h^{1\alpha\beta} \partial^\gamma  h^{4}_{\alpha\beta} +\dots$  \\\hline
$A_2 A_7 B_{12} B_{14} B_{34}+\dots$	&	$\partial^\gamma h^{1\mu\nu}\partial^\beta h^{4\alpha}_\mu h^{2}_{\nu\gamma} h^3_{\alpha\beta}+\rm \dots$	&	$A_2 A_7 B_{13} B_{14} B_{24}+\dots$	&	$ \partial^\beta h^{1\mu\nu} \partial^\gamma h^{4\alpha}_\mu h^{3}_{\nu\gamma} h^2_{\alpha\beta}+\rm \dots$ \\ \hline
$A_1 A_8 B_{23}^2 B_{14}+\dots$	&	$h^{1\mu\nu} h^4_{\mu\gamma}\partial_\nu \partial^\gamma h^{2\alpha\beta}     h^{3}_{\alpha\beta} +\dots$	&	$A_1 A_8 B_{12} B_{23} B_{34}+\dots$	&	$ h^{1\mu\nu}  \partial^\alpha \partial_\nu h^{2}_{\mu\gamma}   h^{3\gamma\beta} h^{4}_{\beta\alpha}+ \dots$ \\ \hline
$A_1 A_8 B_{13} B_{23} B_{24}+\dots$	&	$ h^{1\mu\nu}  \partial^\alpha \partial_\nu h^{3}_{\mu\gamma}   h^{2\gamma\beta} h^{4}_{\beta\alpha}+ \dots$ 	&	$ A_3 A_6 B_{23} B_{14}^2+\dots$	&	$h^{2\mu\nu} h^3_{\mu\gamma} \partial^\gamma h^{1\alpha\beta}    \partial_\nu h^{4}_{\alpha\beta} +\dots$ \\ \hline
$A_3 A_6 B_{12} B_{14} B_{34}+\dots$	&	 $\partial^\beta h^{1\mu\nu} \partial^\gamma h^{4\alpha}_\mu h^{2}_{\nu\gamma} h^3_{\alpha\beta}+\rm \dots$ 	&	$A_3 A_6 B_{13} B_{14} B_{24}+\dots$	&	 $\partial^\gamma h^{1\mu\nu} \partial^\beta h^{4\alpha}_\mu h^{3}_{\nu\gamma} h^2_{\alpha\beta}+\rm \dots$
\end{tabular}
\caption{\label{tab:cont2basis} Explicit basis for the parity-even tensor structures invariant under kinetic permutations and with two derivatives }
\end{table}

\clearpage

\bibliographystyle{JHEP2015}
\bibliography{papers}

\providecommand{\href}[2]{#2}\begingroup\raggedright\begin{thebibliography}{10}

\bibitem{Palti:2019pca}
E.~Palti, \emph{{The Swampland: Introduction and Review}},
  \href{https://doi.org/10.1002/prop.201900037}{\emph{Fortsch. Phys.}
  {\bfseries 67} (2019) 1900037}
  [\href{https://arxiv.org/abs/1903.06239}{{\ttfamily 1903.06239}}].

\bibitem{vanBeest:2021lhn}
M.~van Beest, J.~Calder\'on-Infante, D.~Mirfendereski and I.~Valenzuela,
  \emph{{Lectures on the Swampland Program in String Compactifications}},
  \href{https://doi.org/10.1016/j.physrep.2022.09.002}{\emph{Phys. Rept.}
  {\bfseries 989} (2022) 1} [\href{https://arxiv.org/abs/2102.01111}{{\ttfamily
  2102.01111}}].

\bibitem{Klaewer:2018yxi}
D.~Klaewer, D.~L\"ust and E.~Palti, \emph{{A Spin-2 Conjecture on the
  Swampland}}, \href{https://doi.org/10.1002/prop.201800102}{\emph{Fortsch.
  Phys.} {\bfseries 67} (2019) 1800102}
  [\href{https://arxiv.org/abs/1811.07908}{{\ttfamily 1811.07908}}].

\bibitem{Chowdhury:2019kaq}
S.~D. Chowdhury, A.~Gadde, T.~Gopalka, I.~Halder, L.~Janagal and S.~Minwalla,
  \emph{{Classifying and constraining local four photon and four graviton
  S-matrices}}, \href{https://doi.org/10.1007/JHEP02(2020)114}{\emph{JHEP}
  {\bfseries 02} (2020) 114}
  [\href{https://arxiv.org/abs/1910.14392}{{\ttfamily 1910.14392}}].

\bibitem{Haring:2022cyf}
K.~H\"aring and A.~Zhiboedov, \emph{{Gravitational Regge bounds}},
  \href{https://arxiv.org/abs/2202.08280}{{\ttfamily 2202.08280}}.

\bibitem{deRham:2022hpx}
C.~de~Rham, S.~Kundu, M.~Reece, A.~J. Tolley and S.-Y. Zhou, \emph{{Snowmass
  White Paper: UV Constraints on IR Physics}},  in \emph{{Snowmass 2021}}, 3,
  2022, \href{https://arxiv.org/abs/2203.06805}{{\ttfamily 2203.06805}}.

\bibitem{Chandorkar:2021viw}
D.~Chandorkar, S.~D. Chowdhury, S.~Kundu and S.~Minwalla, \emph{{Bounds on
  Regge growth of flat space scattering from bounds on chaos}},
  \href{https://doi.org/10.1007/JHEP05(2021)143}{\emph{JHEP} {\bfseries 05}
  (2021) 143} [\href{https://arxiv.org/abs/2102.03122}{{\ttfamily
  2102.03122}}].

\bibitem{deRham:2022gfe}
C.~de~Rham, S.~Jaitly and A.~J. Tolley, \emph{{Constraints on Regge behavior
  from IR physics}},
  \href{https://doi.org/10.1103/PhysRevD.108.046011}{\emph{Phys. Rev. D}
  {\bfseries 108} (2023) 046011}
  [\href{https://arxiv.org/abs/2212.04975}{{\ttfamily 2212.04975}}].

\bibitem{Noumi:2022wwf}
T.~Noumi and J.~Tokuda, \emph{{Finite energy sum rules for gravitational Regge
  amplitudes}}, \href{https://doi.org/10.1007/JHEP06(2023)032}{\emph{JHEP}
  {\bfseries 06} (2023) 032}
  [\href{https://arxiv.org/abs/2212.08001}{{\ttfamily 2212.08001}}].

\bibitem{Hamada:2023cyt}
Y.~Hamada, R.~Kuramochi, G.~J. Loges and S.~Nakajima, \emph{{On (scalar QED)
  gravitational positivity bounds}},
  \href{https://doi.org/10.1007/JHEP05(2023)076}{\emph{JHEP} {\bfseries 05}
  (2023) 076} [\href{https://arxiv.org/abs/2301.01999}{{\ttfamily
  2301.01999}}].

\bibitem{Camanho:2014apa}
X.~O. Camanho, J.~D. Edelstein, J.~Maldacena and A.~Zhiboedov, \emph{{Causality
  Constraints on Corrections to the Graviton Three-Point Coupling}},
  \href{https://doi.org/10.1007/JHEP02(2016)020}{\emph{JHEP} {\bfseries 02}
  (2016) 020} [\href{https://arxiv.org/abs/1407.5597}{{\ttfamily 1407.5597}}].

\bibitem{Maldacena:2015waa}
J.~Maldacena, S.~H. Shenker and D.~Stanford, \emph{{A bound on chaos}},
  \href{https://doi.org/10.1007/JHEP08(2016)106}{\emph{JHEP} {\bfseries 08}
  (2016) 106} [\href{https://arxiv.org/abs/1503.01409}{{\ttfamily
  1503.01409}}].

\bibitem{Chakraborty:2020rxf}
S.~Chakraborty, S.~D. Chowdhury, T.~Gopalka, S.~Kundu, S.~Minwalla and
  A.~Mishra, \emph{{Classification of all 3 particle S-matrices quadratic in
  photons or gravitons}},
  \href{https://doi.org/10.1007/JHEP04(2020)110}{\emph{JHEP} {\bfseries 04}
  (2020) 110} [\href{https://arxiv.org/abs/2001.07117}{{\ttfamily
  2001.07117}}].

\bibitem{Bonifacio:2018vzv}
J.~Bonifacio and K.~Hinterbichler, \emph{{Bounds on Amplitudes in Effective
  Theories with Massive Spinning Particles}},
  \href{https://doi.org/10.1103/PhysRevD.98.045003}{\emph{Phys. Rev. D}
  {\bfseries 98} (2018) 045003}
  [\href{https://arxiv.org/abs/1804.08686}{{\ttfamily 1804.08686}}].

\bibitem{Bonifacio:2018aon}
J.~Bonifacio and K.~Hinterbichler, \emph{{Universal bound on the strong
  coupling scale of a gravitationally coupled massive spin-2 particle}},
  \href{https://doi.org/10.1103/PhysRevD.98.085006}{\emph{Phys. Rev. D}
  {\bfseries 98} (2018) 085006}
  [\href{https://arxiv.org/abs/1806.10607}{{\ttfamily 1806.10607}}].

\bibitem{Bonifacio:2019mgk}
J.~Bonifacio, K.~Hinterbichler and R.~A. Rosen, \emph{{Constraints on a
  gravitational Higgs mechanism}},
  \href{https://doi.org/10.1103/PhysRevD.100.084017}{\emph{Phys. Rev. D}
  {\bfseries 100} (2019) 084017}
  [\href{https://arxiv.org/abs/1903.09643}{{\ttfamily 1903.09643}}].

\bibitem{Bonifacio:2019ioc}
J.~Bonifacio and K.~Hinterbichler, \emph{{Unitarization from Geometry}},
  \href{https://doi.org/10.1007/JHEP12(2019)165}{\emph{JHEP} {\bfseries 12}
  (2019) 165} [\href{https://arxiv.org/abs/1910.04767}{{\ttfamily
  1910.04767}}].

\bibitem{Hinterbichler:2017qyt}
K.~Hinterbichler, A.~Joyce and R.~A. Rosen, \emph{{Massive Spin-2 Scattering
  and Asymptotic Superluminality}},
  \href{https://doi.org/10.1007/JHEP03(2018)051}{\emph{JHEP} {\bfseries 03}
  (2018) 051} [\href{https://arxiv.org/abs/1708.05716}{{\ttfamily
  1708.05716}}].

\bibitem{Bonifacio:2017nnt}
J.~Bonifacio, K.~Hinterbichler, A.~Joyce and R.~A. Rosen, \emph{{Massive and
  Massless Spin-2 Scattering and Asymptotic Superluminality}},
  \href{https://doi.org/10.1007/JHEP06(2018)075}{\emph{JHEP} {\bfseries 06}
  (2018) 075} [\href{https://arxiv.org/abs/1712.10020}{{\ttfamily
  1712.10020}}].

\bibitem{deRham:2014zqa}
C.~de~Rham, \emph{{Massive Gravity}},
  \href{https://doi.org/10.12942/lrr-2014-7}{\emph{Living Rev. Rel.} {\bfseries
  17} (2014) 7} [\href{https://arxiv.org/abs/1401.4173}{{\ttfamily
  1401.4173}}].

\bibitem{Hinterbichler:2011tt}
K.~Hinterbichler, \emph{{Theoretical Aspects of Massive Gravity}},
  \href{https://doi.org/10.1103/RevModPhys.84.671}{\emph{Rev. Mod. Phys.}
  {\bfseries 84} (2012) 671} [\href{https://arxiv.org/abs/1105.3735}{{\ttfamily
  1105.3735}}].

\bibitem{Bonifacio:2016wcb}
J.~Bonifacio, K.~Hinterbichler and R.~A. Rosen, \emph{{Positivity constraints
  for pseudolinear massive spin-2 and vector Galileons}},
  \href{https://doi.org/10.1103/PhysRevD.94.104001}{\emph{Phys. Rev. D}
  {\bfseries 94} (2016) 104001}
  [\href{https://arxiv.org/abs/1607.06084}{{\ttfamily 1607.06084}}].

\bibitem{deRham:2017xox}
C.~de~Rham, S.~Melville and A.~J. Tolley, \emph{{Improved Positivity Bounds and
  Massive Gravity}}, \href{https://doi.org/10.1007/JHEP04(2018)083}{\emph{JHEP}
  {\bfseries 04} (2018) 083}
  [\href{https://arxiv.org/abs/1710.09611}{{\ttfamily 1710.09611}}].

\bibitem{Bellazzini:2017fep}
B.~Bellazzini, F.~Riva, J.~Serra and F.~Sgarlata, \emph{{Beyond Positivity
  Bounds and the Fate of Massive Gravity}},
  \href{https://doi.org/10.1103/PhysRevLett.120.161101}{\emph{Phys. Rev. Lett.}
  {\bfseries 120} (2018) 161101}
  [\href{https://arxiv.org/abs/1710.02539}{{\ttfamily 1710.02539}}].

\bibitem{deRham:2018qqo}
C.~de~Rham, S.~Melville, A.~J. Tolley and S.-Y. Zhou, \emph{{Positivity Bounds
  for Massive Spin-1 and Spin-2 Fields}},
  \href{https://doi.org/10.1007/JHEP03(2019)182}{\emph{JHEP} {\bfseries 03}
  (2019) 182} [\href{https://arxiv.org/abs/1804.10624}{{\ttfamily
  1804.10624}}].

\bibitem{Alberte:2019xfh}
L.~Alberte, C.~de~Rham, A.~Momeni, J.~Rumbutis and A.~J. Tolley,
  \emph{{Positivity Constraints on Interacting Spin-2 Fields}},
  \href{https://doi.org/10.1007/JHEP03(2020)097}{\emph{JHEP} {\bfseries 03}
  (2020) 097} [\href{https://arxiv.org/abs/1910.11799}{{\ttfamily
  1910.11799}}].

\bibitem{Wang:2020xlt}
Z.-Y. Wang, C.~Zhang and S.-Y. Zhou, \emph{{Generalized elastic positivity
  bounds on interacting massive spin-2 theories}},
  \href{https://doi.org/10.1007/JHEP04(2021)217}{\emph{JHEP} {\bfseries 04}
  (2021) 217} [\href{https://arxiv.org/abs/2011.05190}{{\ttfamily
  2011.05190}}].

\bibitem{Bellazzini:2023nqj}
B.~Bellazzini, G.~Isabella, S.~Ricossa and F.~Riva, \emph{{Massive Gravity is
  not Positive}},  \href{https://arxiv.org/abs/2304.02550}{{\ttfamily
  2304.02550}}.

\bibitem{Costa:2011mg}
M.~S. Costa, J.~Penedones, D.~Poland and S.~Rychkov, \emph{{Spinning Conformal
  Correlators}}, \href{https://doi.org/10.1007/JHEP11(2011)071}{\emph{JHEP}
  {\bfseries 11} (2011) 071} [\href{https://arxiv.org/abs/1107.3554}{{\ttfamily
  1107.3554}}].

\bibitem{Bonifacio:2017iry}
J.~J. Bonifacio, \emph{{Aspects of Massive Spin-2 Effective Field Theories}},
  Ph.D. thesis, Oxford U., 2017.

\bibitem{deRham:2017zjm}
C.~de~Rham, S.~Melville, A.~J. Tolley and S.-Y. Zhou, \emph{{UV complete me:
  Positivity Bounds for Particles with Spin}},
  \href{https://doi.org/10.1007/JHEP03(2018)011}{\emph{JHEP} {\bfseries 03}
  (2018) 011} [\href{https://arxiv.org/abs/1706.02712}{{\ttfamily
  1706.02712}}].

\bibitem{osti_4128791}
A.~Kotanski, \emph{Transversity amplitudes and their application to the study
  of collisions of particles with spin.}, {\emph{Acta Phys. Pol. B1:
  45-58(1970).} (1970) }.

\bibitem{Kravchuk:2016qvl}
P.~Kravchuk and D.~Simmons-Duffin, \emph{{Counting Conformal Correlators}},
  \href{https://doi.org/10.1007/JHEP02(2018)096}{\emph{JHEP} {\bfseries 02}
  (2018) 096} [\href{https://arxiv.org/abs/1612.08987}{{\ttfamily
  1612.08987}}].

\bibitem{Cohen-Tannoudji:1968lnm}
G.~Cohen-Tannoudji, A.~Morel and H.~Navelet, \emph{{Kinematical singularities,
  crossing matrix and kinematical constraints for two-body helicity
  amplitudes}},
  \href{https://doi.org/10.1016/0003-4916(68)90243-1}{\emph{Annals Phys.}
  {\bfseries 46} (1968) 239}.

\bibitem{kota}
A.~Kotański, \emph{{Kinematical singularities of the transversity
  amplitudes}},
  \href{https://doi.org/https://doi.org/10.1007/BF02819831}{\emph{Il Nuovo
  Cimento A (1965-1970)} {\bfseries 56} (1968) 239}.

\bibitem{Kotanski:1965zz}
A.~Kotanski, \emph{{DIAGONALIZATION OF HELICITY CROSSING MATRICES}}, .

\end{thebibliography}\endgroup

\end{document}